\documentclass{sig-alternate-05-2015}

\usepackage{graphicx}

\makeatletter
\def\@copyrightspace{\relax}
\makeatother

\begin{document}






\title{Application of Advanced Record Linkage Techniques \\
       for Complex Population Reconstruction}

%
%
%
%
%

\numberofauthors{1} 
%
\author{
%
%
\alignauthor
  Peter Christen \\
  \affaddr{Research School of Computer Science} \\
  \affaddr{The Australian National University} \\
  \affaddr{Acton, ACT, 2601, Australia}\\
  \email{peter.christen@anu.edu.au}
}

\newtheorem{definition}{Definition}


\maketitle

\begin{abstract}
Record linkage is the process of identifying records that refer to
the same entities from several databases. This process is challenging
because commonly no unique entity identifiers are available. Linkage
therefore has to rely on partially identifying attributes, such as
names and addresses of people. Recent years have seen the development
of novel techniques for linking data from diverse application areas,
where a major focus has been on linking complex data that contain
records about different types of entities. Advanced approaches that
exploit both the similarities between record attributes as well as the
relationships between entities to identify clusters of matching
records have been developed.

In this application paper we study the novel problem where rather than
different types of entities we have databases where the same entity
can have different roles, and where these roles change over time. We
specifically develop novel techniques for linking historical birth,
death, marriage and census records with the aim to reconstruct the
population covered by these records over a period of several decades.
Our experimental evaluation on real Scottish data shows that even with
advanced linkage techniques that consider group, relationship, and
temporal aspects it is challenging to achieve high quality linkage
from such complex data.
\end{abstract}





\section{Introduction}
\label{sec-intro}

Record linkage, also known as entity resolution, data matching or
duplicate detection~\cite{Chr12,Her07}, is the process of
identifying and linking records about the same real-world entity from
one or several databases. This process is widely used in the data
preparation phase of data mining projects that require data from
various sources to be integrated before they can be analyzed.
Traditionally, record linkage has been employed in the health sector
and for national censuses, while more recently it has seen application
in a wide range of areas including e-commerce, the social sciences,
crime and fraud detection, and national
security~\cite{Chr12,Her07,Kum14}.

Much research in record linkage is based on bibliographic data (with
the aim to identify which publications have been written by the same
authors) due to the easy availability of large bibliographic databases
when compared to personal data about people (such as health or
government records) that are often sensitive or
confidential~\cite{Kop10b}.

Bibliographic data also have the advantage of containing several types
of entities (like authors, publications, and venues). These provide a
complex information space~\cite{Don05} that allows graph-based and
clustering techniques to exploit both attribute similarities (such as
similar publication titles) as well as relationships (for example two
authors have co-written several publications). Several advanced graph
and collective linkage techniques have been developed for
bibliographic data in recent years~\cite{Bha07,Don05}

The question of how to employ such advanced linkage techniques on
personal data (such as names, addresses, dates of birth, etc.) has
however so far only seen very limited research~\cite{Fu14b,Fu12}.
A major reason for this is the lack of available real-world databases
about people.
Most approaches used to link personal data are still based on
traditional techniques that only consider pair-wise record
comparisons~\cite{Her07}. These techniques classify each record pair
individually as either a \emph{match} (assumed to refer to the same
entity) or a \emph{non-match} (assumed to refer to different entities)
without considering relationships between records.

Recent years have seen a significant increase in interest on how to
efficiently and effectively link large databases about people, with
the aim to reconstruct populations for social science and health
research, or for use as a valuable resource for governments and
businesses~\cite{Ant14a,Blo15}. The emerging field of \emph{population
informatics}~\cite{Kum14} is aimed at the linkage and analysis of
large and dynamic databases that contain information about
people, such as the health, education, financial, census, location,
shopping, employment, and social networking records of a large
proportion of individuals in a population.

Compared to bibliographic data with their different types of entities,
population databases contain records about people where the role of an
individual can change over time, and where different databases contain
different types of information about the same individual. Furthermore,
the different roles of individuals lead to various types of
relationships and diverse constraints (with regard to roles, time, and
number of individuals involved in a link).

\begin{figure}[t!]
  \centering
  \includegraphics[width=0.48\textwidth]
                  {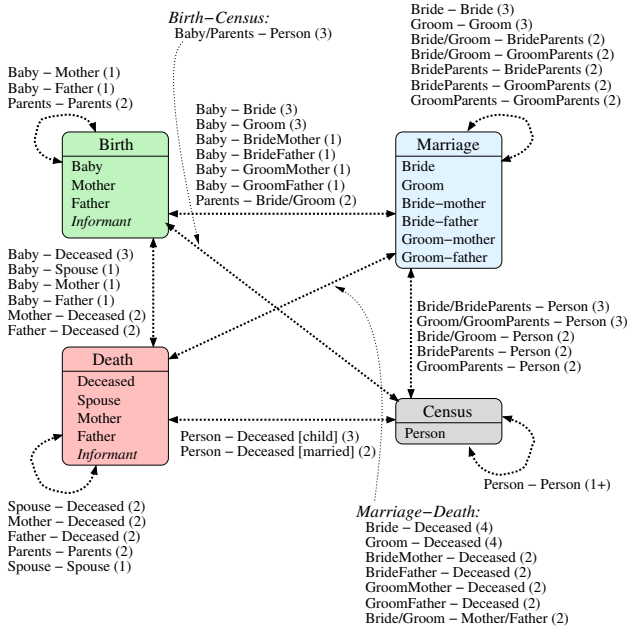}
  \caption{Structure of the domain of birth, death, marriage and
           census certificates we consider in this paper. For each
           role pair, the given numbers show how many individuals can
           be linked between certificates of such a pair. The outcomes
           of linking certificates can then be converted into life
           segments such as the ones shown in Fig.~\ref{fig-segment}.
           \label{fig-bdmc}}
\end{figure}

This is illustrated in Fig.~\ref{fig-bdmc}, which shows the roles and
relationships in the data sets we use in our evaluation in
Sect.~\ref{sec-data}. For each of the pairs of roles shown (for
example \emph{Baby--Bride}), different attributes (fields) are
available for calculating the similarities between records, different
temporal constraints are possible (for example, a \emph{Baby} can only
become a \emph{Bride} after a certain age), and different 1-to-1 or
1-to-many linkage restrictions need to be considered (a \emph{Baby}
can be a \emph{Bride} in one or more marriages, but each \emph{Bride}
can only have been a \emph{Baby} once).

\begin{figure}[t!]
  \centering
  \includegraphics[width=0.47\textwidth]
                  {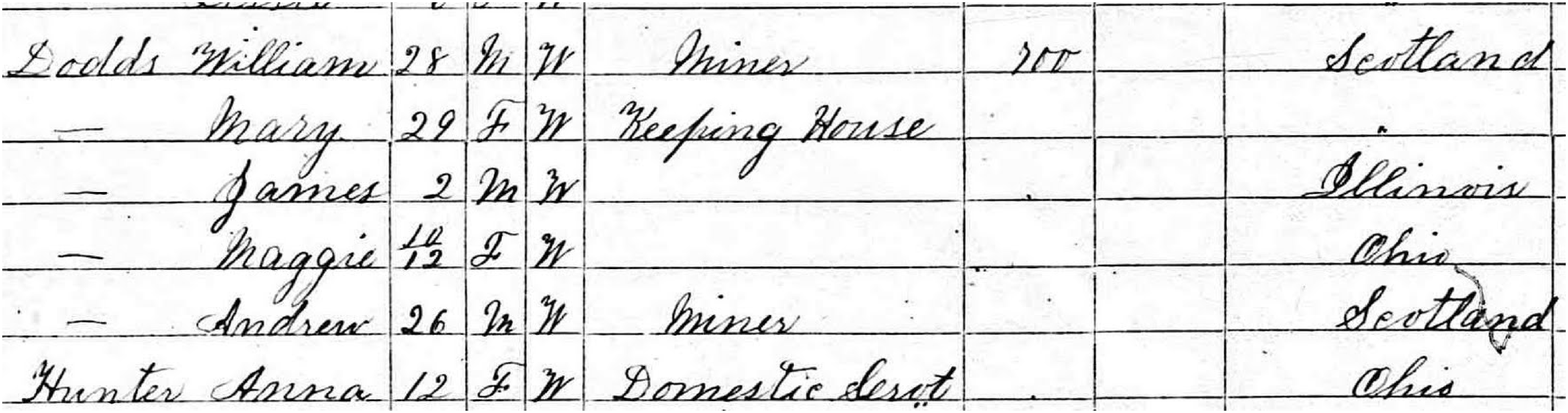}
  \caption{Small sample of a historical census certificate showing one
           family and their servant. \label{fig-census}}
\end{figure}

The linkage of such data is challenging due to people moving around
(address changes), changing their names (either in the form of small
spelling variations or errors, or full name changes due to marriage),
and other data quality issues such as transcription or scanning errors
when these certificates were converted from their original handwritten
into digital form (as illustrated in Fig.~\ref{fig-census}). People in
the 19th century also did not always know their exact date of birth or
even their true age, leading to inaccurate age values in certificates.
Additionally, the variety of personal names was much more limited (as
Table~\ref{table-data} shows for our data sets), and often only a few
common names covered a large proportion of a population. This means
that, compared to many modern databases, the use of attribute
similarities does not provide enough information to distinguish
between individuals.

On the other hand, the various types of relationships can help improve
linkage quality. For example, a \emph{Baby} and her parents occurring
in a census certificate (as a child and her parents) that have a high
similarity to a \emph{Bride} and her parents in a marriage
certificate, will provide supportive evidence that the \emph{Baby} and
\emph{Bride} refer to the same person.

\begin{figure*}[t!]
  \centering
  \includegraphics[width=1.0\textwidth]
                  {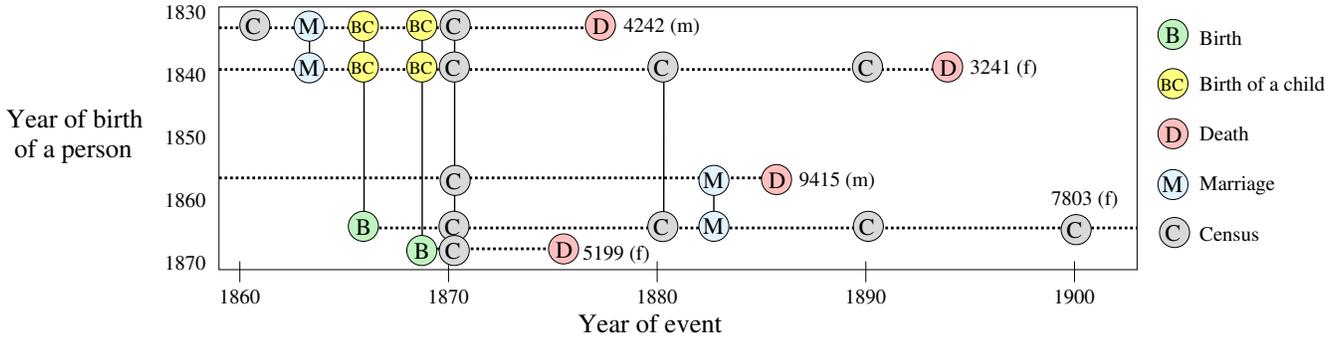}
  \caption{Example life segments for a couple (husband with person
           identifier `4242' and wife with identifier `3241'), their
           daughters `7803' and `5199', and the husband `9415' of
           daughter `7803'. Each set of nodes linked via a solid
           vertical line corresponds to a certificate such as those
           shown in Fig.~\ref{fig-bdmc} (they show how several
           individuals occur together on the same certificate),
           while the dotted horizontal lines represent the events for
           one individual. These horizontal lines between certificates
           are the links we aim to identify in our work.
           \label{fig-segment}}
\end{figure*}

The outcomes of linking such population databases are linked records
for each individual that can then be visualized into \emph{life
segments} (as shown in Fig.~\ref{fig-segment}) or converted into
\emph{family trees}, and used for example in health~\cite{Kum14} and
genealogical research studies~\cite{Blo15}.

While most techniques for reconstructing
populations~\cite{Ant14a,Blo15,Fu14b,Fu12}, including those in the
present paper, focus on historical data (due to their easier
availability compared to contemporary data) they are also applicable
to modern data that have similar structure and content.
\smallskip

\textbf{Contributions:}
While graph and collective linkage techniques~\cite{Bha07,Don05} use
both attribute and relational similarities between entities of
\emph{different types} (such as papers, authors and venues in
publications), we study the problem of linking records about entities
(such as people) that are \emph{of the same type} but can have
\emph{different roles} in different records, where these roles can
\emph{change over time}, and where there are \emph{constraints} both
for the roles and relationships of an entity.

We specifically investigate how advanced record linkage techniques can
be successfully applied for the reconstruction of complex populations
such as those collected by governments and health organizations. We
develop a novel linkage approach that combines pair-wise attribute
and group similarities, incorporates temporal as well as 1-to-1 and
1-to-many linkage constraints, and considers relationship information
between individuals and groups of people as for example occur on
certificates such as those shown in Fig.~\ref{fig-bdmc}. We show that
even state-of-the-art advanced linkage techniques have difficulty to
achieve high linkage quality in an experimental evaluation on a real
collection of Scottish data sets that span over five decades of the
nineteenth century.

\section{Related Work}
\label{sec-related}

The challenges of how to link records about the same entities within
and across databases have been investigated in various research
domains since the 1950s. Several recent books provide an overview of
the topic~\cite{Chr12,Her07,Nau10}.
In the following we describe related work specifically aimed at
linking complex data. These approaches include group, graph and
collective techniques that consider data that contain temporal
aspects, or that are specific to linking (historical) census and
similar types of personal data.

Traditional record linkage techniques only consider the similarities
between record attributes mainly using approximate string comparison
functions~\cite{Chr12,Her07}, followed by the individual
classification of each compared record pair as a match or a non-match.
On et al.~\cite{On07} were the first to investigate how to link groups
of records in the context of bibliographic data (such as groups of
papers written by the same author) using a weighted bipartite matching
approach. While this approach still classifies each pair of groups of
records independently from all others, recently developed collective
techniques~\cite{Bha07,Don05} additionally also consider the
relationships between entities. They build a graph of all compared
records that is clustered such that each cluster refers to one entity.
For bibliographic data, the available relationships include authors
who have co-authored together, or who work at the same institution.

These group and graph-based techniques have been developed in the
context of bibliographic data, where several types of entities are
available. Fu et al.~\cite{Fu12} were the first to investigate how
group linkage techniques can be employed in the context of
(historical) census data by using household information to group
records with the aim of linking individuals by first linking similar
households. Experimental results showed that this approach can lead
to a significant reduction of ambiguous links between individuals.

While most record linkage techniques do not consider temporal aspects,
Li et al.~\cite{Li11b} and Chiang et al.~\cite{Chi14} investigated
temporal information in the context of bibliographic data where
authors can change their affiliations over time. Their approach
adjusts similarity weights depending upon the temporal differences
between two records. Christen and Gayler~\cite{Chr13} developed a
similar approach for personal data where temporal likelihood values,
such as how likely it is that somebody moves address over a certain
period of time, are learned from a large population database. In our
approach we also consider temporal relationships between records, for
example that a birth has to occur before any other records of the
same person, or that a marriage or birth of a child can only occur
once a person has reached a certain age.

To improve linkage quality, various supervised learning techniques
have been developed specifically aimed at linking (historical) census
data. Fu et al.~\cite{Fu12} investigated multiple instance learning by
first linking individuals with high confidence, and then using those
to link households over time. More recently, the same
authors~\cite{Fu14b} exploited information in households that does not
change over time (such as the age differences between a husband and
wife) within a graph-matching approach to improve the quality of
linking census data over time. Antoine et al.~\cite{Ant14a} used a
support vector machine classifier to successfully link two large
Canadian census data sets from 1871 and 1881.

The main drawback of supervised approaches is their reliance on
training data, which generally have to be manually prepared by domain
experts in an expensive and time consuming process. Our approach, on
the other hand, is unsupervised and aims to exploit the relationships
between individuals and their changing roles over time to achieve
high linkage quality.



\begin{figure*}[t!]
  \centering
  \includegraphics[width=0.99\textwidth]
                  {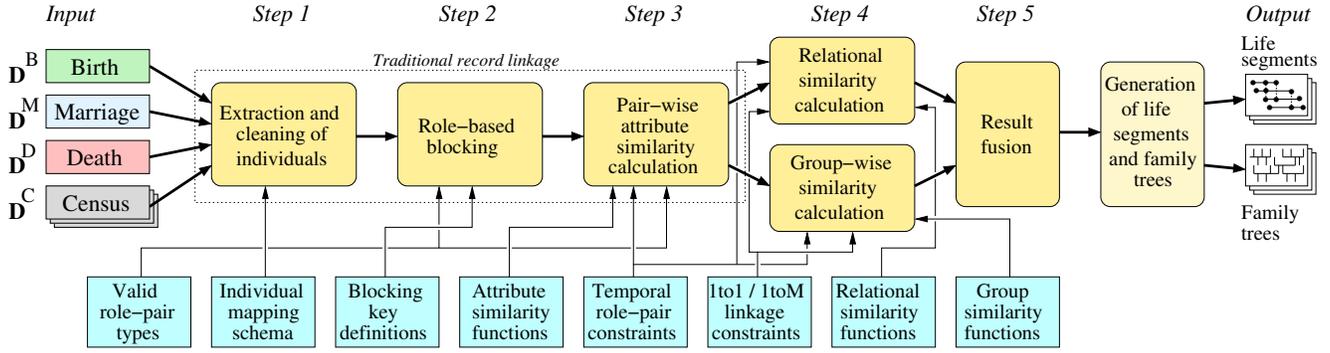}
  \caption{Pipeline of the main steps involved in our approach to
           complex population record linkage, as described in
           Sect.~\ref{sec-poplink}. \label{fig-pipeline}}
\end{figure*}

\section{Problem Definition}
\label{sec-def}

We now define the problem we aim to tackle using the following
notation. We assume a collection of data sets that contain details
about entities (in our case people), where each data set contains
\emph{certificates} of different types. We assume birth
($\mathbf{D}^B$), death ($\mathbf{D}^D$), marriage ($\mathbf{D}^M$),
and census ($\mathbf{D}^C$) data sets. These four types are common
both in historical as well as contemporary data as collected for
example by government registries and census agencies~\cite{Blo15}. We
denote certificates as $C_i^B \in \mathbf{D}^B$, $C_i^D \in
\mathbf{D}^D$, $C_i^M \in \mathbf{D}^M$, and $C_i^C \in \mathbf{D}^C$.
We drop the superscript ($^B$, $^C$, etc.) if a data set or
certificate can be of any type. Each certificate contains the details
of one or several \emph{individuals}, as shown in Fig.~\ref{fig-bdmc}.
Each certificate also has an event year, $C_i.y$, such as the year of
a birth, a marriage, or a death, or the year a census was conducted,
as illustrated in Fig.~\ref{fig-segment}.

A \emph{record} for an individual in a certificate is denoted as
$r_{i,j} \in C_i$, where $\mathbf{R}$ is the set of all individual
records from all data sets, i.e.\ $\mathbf{R} = \{r_{i,j} \in C_i :
C_i \in  \mathbf{D}^B \cup \mathbf{D}^D \cup \mathbf{D}^M \cup
\mathbf{D}^C\}$. A record contains a set of attributes $A$, where
$r_{i,j}.a$ is the value of attribute $a \in A$ for record $r_{i,j}$.
Each record represents an entity (person), and we use $r_{i,j}.e$
to denote the entity identifier for that record. Two records that have
the same value in their entity identifiers are a \emph{match}, i.e.\
they are assumed to correspond to the same person. It is important to
note that all individual records in a certificate are about different
people, i.e.\ $r_{i,j}.e \ne r_{i,k}.e ~\forall~ j \ne k \land r_{i,j},
r_{i,k} \in C_i$. We denote with $\mathbf{E}$ the set of all entities
(individuals) that occur in all records in $\mathbf{R}$, i.e.\
$\mathbf{E} = \{r_{i,j}.e: ~\forall~ r_{i,j} \in \mathbf{R}\}$.

We assume a set of \emph{role types} $\mathbf{T} = \mathbf{T}^B \cup
\mathbf{T}^D \cup \mathbf{T}^M \cup \mathbf{T}^C$, such as those shown
in Fig.~\ref{fig-bdmc}, where different data sets contain records of
individuals with different roles. For example, $\mathbf{T}^B = \{
Baby, Mother, Father, Informant\}$. Each individual record has one
such role type, denoted as $r_{i,j}.t$. From these roles, we are
interested in pairs of roles that correspond to the roles a person can
have in the different pairs of certificates. As an example, the pair
$(Baby,Bride)$ corresponds to a baby in a birth certificate who
becomes the bride in a marriage certificate. We denote the set of all
valid role pair types as $\mathbf{P} = \{(t_i,t_j) : t_i, t_j \in
\mathbf{T}\}$. Note that not all possible pairs that could be
generated from roles in $\mathbf{T}$ are in $\mathbf{P}$. For example
$(Groom,Mother) \notin \mathbf{P}$ as this is not a valid pair of
roles for the same individual. Limiting role pairs to those in
$\mathbf{P}$ ensures that we do not link pairs of records with invalid
role pairs. This helps both to reduce the number of records to
compare, as well as to improve linkage quality, as we discuss in the
following section.

\newpage

Finally, we define a life segment $\mathbf{l}$ as a sequence of one or
more individual records $\mathbf{l} = [r_{i_{1},a},r_{i_{2},b}, \ldots,
r_{i_{n},c}]$, with $n = |\mathbf{l}|$, ordered according to the event
years of their corresponding certificates, i.e.\ $C_{i_1}.y \le
C_{i_2}.y \le \ldots \le C_{i_n}.y$, where the role types of all pairs
of records in a life segment are valid role pairs, i.e.\
$(r_{i,a}.t,r_{j,b}.t) \in \mathbf{P} ~\forall~ r_{i,a}, r_{j,b} \in
\mathbf{l}$. These life segments can be converted into family trees as
used in genealogical studies~\cite{Blo15} (this process is outside of
the topic of this paper). We now formally define the problem we aim to 
address in this paper:
\medskip

\emph{Definition 1 [Linkage-based complex population reconstruction]:}
Given data sets $\mathbf{D}^B$, $\mathbf{D}^D$, $\mathbf{D}^M$, and
$\mathbf{D}^C$ with different types of certificates, where each
certificate contains one or more records about individual entities
$e_k \in \mathbf{E}$, the aim of linkage-based complex population
reconstruction is to link certificates and individual records to
generate a set $\mathbf{L}$ of life segments such that for each entity
$e_k \in \mathbf{E}$ there will be one life segment $\mathbf{l}_k \in
\mathbf{L}$ that only contains records about entity $e_k$, and no
other life segment in $\mathbf{L}$ contains any records of entity
$e_k$, i.e.\ $r_{i,j}.e = e_k ~\forall~ r_{i,j} \in \mathbf{l}_k \land
r_{x,y}.e \ne e_k ~\forall~ r_{x,y} \in \mathbf{l}_z: \mathbf{l}_k \ne
\mathbf{l}_z, ~\forall~ \mathbf{l}_k, \mathbf{l}_z \in \mathbf{L}$.
\medskip

The following questions now arise: how can we (1) calculate
similarities between pairs and groups of records and certificates;
(2) take relationships, roles, and temporal aspects into account
during this linkage process; and (3) conduct the linkage efficiently
and effectively to generate life segments from potentially very
large population databases.


\section{Complex Population Linkage}
\label{sec-poplink}

In this section we discuss all steps involved in our approach to
linking complex population databases, as illustrated in
Fig.~\ref{fig-pipeline}. We first discuss the three steps involved in
the traditional pair-wise linkage of individuals, then describe the
use of temporal, and 1-to-1 and 1-to-many constraints, then provide
details of the advanced relational and group linkage approaches, and
finally discuss the result fusion step.


\subsection{Pair-wise Linkage of Individuals}
\label{sec-pair-wise}

The input to the pair-wise linkage steps are the data sets
$\mathbf{D}^B$, $\mathbf{D}^D$, $\mathbf{D}^M$, and $\mathbf{D}^C$,
which contain different types of certificates as described above. We
first extract and clean the attribute values required for linkage for
all individuals from the different certificates~\cite{Chr12}. We then
use a mapping schema which maps each individual record $r_{i,j} \in
\mathbf{R}$ into a single table with a common set of attributes (such
as first and last name, gender, address, role, occupation, and so on),
together with a reference to its source certificate $C_i$ and its
event year $C_i.y$.

In the second step we apply blocking with the aim to extract sets of
records that likely correspond to the same entity. Blocking is
commonly used in record linkage to reduce the number of record pairs
that are to be compared from the full quadratic comparison
space~\cite{Chr12,Her07}. We use a blocking criteria $B$ that consists
of a set of blocking key definitions $b \in B$ to split the data sets
into blocks, such that all records with the same blocking key value
are inserted into the same block. Each blocking key is a tuple
$b=(\mathbf{T}_b,a_x \ldots a_z)$, where $\mathbf{T}_b \subset
\mathbf{T}$ is a set of role types and $a_x, \ldots, a_z \in A$ are
one or more attributes.

We apply phonetic encoding on name and address attribute values to
group variations of similar sounding values into the same block. In
our application we use Double-Metaphone which overcomes some of the
drawbacks of the commonly used Soundex method, such as sensitivity to
different first letters~\cite{Chr12}. All records $r_{i,j}$ with
$r_{i,j}.t \in \textbf{T}_b$ that have the same value in their
concatenated phonetically encoded attribute values $r_{i,j}.a_x,
\ldots, r_{i,j}.a_z$ are inserted into the same block. We only insert
records into the same block if pairs of their role types are in the
set $\mathbf{P}$ of valid role pair types to prevent comparing records
with invalid roles (for example, records with types $Groom$ and
$Bride$ are never inserted into the same block).

Records in the same block are compared pair-wise in the third step
using a set of similarity functions, such as edit distance or
Jaro-Winker~\cite{Her07} for string attributes, and an approximate
year difference function~\cite{Chr12} to allow for variations in year
values.
For each compared record pair, the similarity $sim(r_{i,j},r_{k,l})$ is
then normalized into $[0,1]$, where a similarity of $0$ means two
records are totally different, $1$ means they have exactly the same
values in the compared attributes, and values in-between $0$ and $1$
mean records are somewhat similar.
Only pairs with a similarity $sim(r_{i,j},r_{k,l}) \ge s_m$, where $s_m$
is a minimum similarity threshold, are kept in the set $\mathbf{S}$
of linked pairs along with their similarities, as pairs with lower
similarities likely correspond to different entities. As with
traditional record linkage~\cite{Her07}, this pair-wise linkage step
does not take any group or relationship aspects into account but only
calculates attribute based similarities~\cite{Bha07}. We however 
enforce temporal constraints to prevent definite non-matching pairs
from being compared ($Deceased$ records are not compared to $Baby$
records with later event years).

The set of calculated individual pair-wise similarities, $\mathbf{S}$,
is the input to the two advanced linkage approaches used in step 4.
Each of these two approaches is conducted individually and
incorporates information from linked individuals, role type pairs, the
relationships between individuals, and temporal constraints, with the
aim to identify links of high quality and high confidence between
certificates. The minimum similarity $s_m$, used in the pair-wise
similarity calculation (step 3), is kept at a low value to ensure all
links between matching individuals are included in  $\mathbf{S}$. We
aim for a high recall of true links in $\mathbf{S}$ at the cost of low
precision. The aim of the relational and group-based linkage steps is
to improve precision by identifying the set of true links from the
potentially very large set $\mathbf{S}$ of linked individuals.



\subsection{Temporal Linkage Constraints}
\label{sec-temp-rest}

Temporal linkage constraints can be considered both at the level of
linking records of individuals, as well as at the level of linking
certificates. At the individual level, for each role type pair in the
set $\mathbf{P}$ we define a time interval $[\Delta_{min},$
$\Delta_{max}]$ which determines if two individual records are to be
compared or not in the pair-wise attribute similarity calculation
step. For two individual records $r_{i,j}$ and $r_{k,l}$ from two
certificates $C_i$ and $C_k$ with event years $C_i.y$ and $C_k.y$,
respectively, the pair will be compared if the following equation
evaluates to true:
\begin{equation}
\sigma_{r_{i,j}, r_{k,l}} = \Delta_{min} \le (C_i.y - C_k.y) \le 
                       \Delta_{max},
\end{equation}
assuming $C_i.y \ge C_k.y$. For example, for the role type pair
$(Baby,Deceased)$, $\Delta_{min}=0$ and $\Delta_{max}=999$, indicating
a death record for an individual can only occur after the person's
birth. In our experiments, we however set $\Delta_{min}=-2$ for this
and related role pairs because due to data quality issues it is
possible that some birth and death certificates have wrongly recorded
event years. As a second example, for the role type pair
$(Bride,BrideMother)$, we set $\Delta_{min}=12$ and $\Delta_{max}=999$
assuming a woman can only become a bride's mother once she is old
enough (again providing several years tolerance to account for wrong
event years).

While such temporal constraints limit which pairs of individuals are
compared or not, similar constraints can also be applied at the level
of linking pairs of certificates in the relational and group linkage
steps. At this level, temporal constraints ensure that certificate
links are limited to valid time periods. For example, a marriage and a
census certificate can be linked in different ways. One is to link the
bride and her parents to an earlier census certificate of this family,
while another way is to link the bride and groom to a later census
certificate as a new family.

Our temporal linkage constraints consist of a list of role type pairs
and certificate pairs and their corresponding time intervals
$[\Delta_{min},\Delta_{max}]$. Not all role type pairs and not all
certificate pairs have temporal constraints. For example, a women can
have the roles $(Mother,Mother)$ across multiple birth certificates
spread over several years. 


\subsection{1-to-1 and 1-to-Many Linkage Constraints}
\label{sec-temp-1-1}

Similar to temporal constraints, \emph{1-to-1} and \emph{1-to-many}
(and \emph{many-to-1}) constraints are applicable both at the level
of pairs of individual records, as well as at the level of pairs of
certificates. An example of a 1-to-1 constraint is that a $Baby$ in a
birth record can be a true match to only one $Deceased$ individual in
a death record, and vice versa; while a 1-to-many constraint is where
a $Baby$ can be a true match to a $Bride$ in one or more marriage
certificate (assuming a women re-marries), while each $Bride$ can only
be a true match to a $Baby$ in one birth certificate.

Such 1-to-1 and 1-to-many linkage constraints are often applied in
general record linkage applications where a record in one database can
match to a maximum of one record in a second database~\cite{Chr12}.
The best-matching pairs of records are generally selected based on
their calculated pair-wise similarities, assuming that the record pair
with the highest similarity is the true match. 1-to-1 and 1-to-many
linkage constraints can be implemented in either a greedy fashion or
using an optimal assignment procedure~\cite{Chr12}.

We do not apply such linkage constraints in the pair-wise attribute
similarity step because this might lead to reduced similarities
between certificates in the relational and group linkage steps if
links between individuals are being removed due to 1-to-1 and
1-to-many linkage constraints. For example, if an individual has a low
similarity due to a name variation or wrongly recorded age value it
might not be linked if a 1-to-1 constraint is applied, thereby
lowering the number of individuals that are linked between two
certificates in the group linkage step (as we describe below).

We instead apply 1-to-1 and 1-to-many constraints after the
relational and group linkage steps to reduce the number of linked
certificates further. Our hypothesis is that linked certificates
contain a larger set of evidence (in the group linkage step from
several individuals and in the relational linkage step from the
relationships between individuals) that leads to true matching
certificate pairs with higher similarities compared to non-matching
pairs. Our linkage constraints consist of a list of certificate type
pairs and how they are to be linked, their corresponding
`\emph{1-to-1}', `\emph{1-to-m}', `\emph{m-to-1}' constraints (or
`\emph{m-to-m}' for no constraint), that are applied after the
relational and group linkage steps.


\subsection{Relational Similarity Linkage}
\label{sec-rel-sim}

Based on ideas from graph-based and collective entity resolution
techniques~\cite{Bha07,Don05}, in this step of our approach we
calculate a relational similarity for each pair of certificates that
is based on its neighborhood of certificates. To do so, we first
generate a graph $\mathbf{G}$ where nodes are certificates and edges
between certificates are pairs of individuals from the set
$\mathbf{S}$ that were linked in the pair-wise linkage step and that
had a similarity above the minimum threshold $s_m$. Formally,
$\mathbf{G} = (V,E)$ with the sets of vertices $V = \{\forall~ C_i \in
\mathbf{D}^B \cup \mathbf{D}^D \cup \mathbf{D}^M \cup \mathbf{D}^C\}$
and edges $E = \{\langle C_i,C_k \rangle: sim(r_{i,j}, r_{k,l}) > s_m:
\forall~ r_{i,j} \in C_i \land \forall~ r_{k,l} \in C_k\}$.

For each pair of certificate types (birth to census, birth to death,
etc.) there are generally several ways of how two certificates can be
linked. For example, a birth and marriage certificate can be linked in
the following three ways: (1) a baby girl is linked to her marriage,
(2) a baby boy is linked to his marriage, and (3) a baby's parents are
linked to their marriage. For each such linkage type, we define the
possible relationships with other types of certificates.

\begin{figure}[t!]
  \centering
  \includegraphics[width=0.46\textwidth]
                  {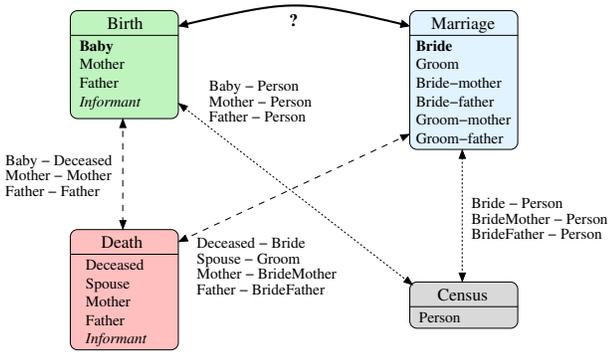}
  \caption{Example relational similarity between a birth and a
           marriage certificate for a baby girl linked to her marriage
           (Bride). As described in Sect.~\ref{sec-rel-sim}, the
           relational similarity between the birth and marriage
           certificate is calculated using the individuals in the
           birth and marriage certificates that are linked to
           neighboring census and death certificates.
           \label{fig-rel}}
\end{figure}

An example is shown in Fig.~\ref{fig-rel} for the linkage of a baby
girl to her marriage ($Bride$), where individual links to neighboring
census and death records in $\mathbf{G}$ are used to calculate the
relational similarity. In this example, if this birth certificate has
several linked individuals with high similarity to the same death and
census certificates as the marriage certificate, then this provides
evidence that they refer to the same group of individuals and should
be linked. On the other hand, if there are no individuals linked to the
same neighboring death and census certificates then this pair of birth
and marriage certificates should not be linked.

We investigate seven relational similarity functions~\cite{Bha07}:
(1) \emph{Jaccard}, which is the size of the intersection of
neighboring certificates divided by the size of the union of
neighbors; (2) \emph{multi-Jaccard}, which also takes the number of
individuals linked between certificates into account; (3) the
\emph{average} individual similarity between common neighboring
certificates without taking the number of linked individuals into
account; (4) \emph{multi-average}, which is the same as (3) but takes
the number of individuals linked between certificates into account;
(5) the \emph{maximum} individual similarity between common
neighboring certificates; (6) the \emph{Adar/Adamic}
similarity~\cite{Bha07} which considers the number of neighbors of
common neighbors (i.e.\ neighbors of neighbors), where a smaller
number of other neighboring certificates means a common neighbor is
more relevant because it is less connected (less
ambiguous)~\cite{Bha07}; and (7) the \emph{multi-Adar/Adamic}
similarity which also takes the number of individuals linked between
certificates into account.

Optional 1-to-1 and 1-to-many linkage constraints can be applied after
the relational similarities between certificates have been calculated
to further reduce the number of links between certificates, and to
ensure no invalid links are being produced (for example, the death of
an individual is not linked to the birth of more than one baby).

\subsection{Group Linkage}
\label{sec-group-sim}

Using group similarities has previously been used in the linkage of
both bibliographic and historical census data~\cite{Fu12,On07}. The
basic idea is to consider all the calculated links between individuals
in a pair of certificates, and based on those to calculate a group
similarity measure between certificates. As with the relational
similarity approach, the input to the group linkage step is the set
$\mathbf{S}$ of all pair-wise linked individuals that have a
similarity of at least $s_m$.

We investigate five group similarity functions: (1) \emph{maximum},
which takes as the group similarity the maximum of the individual
pair-wise similarities for a pair of certificates: $s_{max}(C_i, C_k) =
max(\mathbf{S}_{i,k})$, where $\mathbf{S}_{i,k} = \{sim(r_{i,j},r_{k,l})
\in \mathbf{S}: r_{i,j} \in C_i \land r_{k,l} \in C_k\})$; (2)
\emph{average}, where we calculate the average similarity between all
pairs of individuals for a pair of certificates: $s_{avr}(C_i, C_k) =
(\sum sim(r_{i,j},r_{k,l}) \in \mathbf{S}_{i,k}) / |\mathbf{S}_{i,k}|$;
(3) \emph{group size}, which is the ratio of the number of linked
individuals divided by the larger size of the two certificates,
calculated as $s_{size}(C_i, C_k) = |\mathbf{S}_{i,k}| /
max(|C_i|,|C_k|)$; and (4) \emph{group bipartite}, which follows the
approach of On et al.~\cite{On07} and calculates the similarity
between two certificates as: $s_{grp}(C_i, C_k) = (\sum
sim(r_{i,j},r_{k,l}) \in \mathbf{S}_{i,k}) / (|C_i|+|C_k| -
|\textbf{S}_{i,k}|)$. Finally, our last group similarity function
combines the previous four similarities by calculating their average
as $s_{comb}(C_i, C_k) = (s_{max}(C_i, C_k) + s_{avr}(C_i, C_k) +
s_{size}(C_i, C_k) + s_{grp}(C_i, C_k)) / 4$, with the hypothesis being
that each of these four similarities provides some evidence towards
an overall group similarity for two certificates.

As with the relational similarity approach, optional 1-to-1 and
1-to-many linkage constraints can be applied after the group
similarities between certificates have been calculated to reduce the
number of invalid links.


\subsection{Result Fusion}
\label{sec-fusion}

The final step (step 5 in Fig.~\ref{fig-pipeline}) of our complex
population record linkage approach is to combine the linkage results
obtained from the relational and group linkage steps with the aim to
find true matching certificates. Our assumption is that true matches
should have both a high group similarity (i.e.\ have several
individuals linked between them each with a high similarity) as well
as a high relational similarity (i.e.\ be linked to a similar set of
neighboring certificates).

The input to the fusion step are two sets of linked certificates, 
$\mathbf{M}_R$ (relational matches) and $\mathbf{M}_G$ (group
matches), respectively. Our fusion approach is to calculate a
(weighted) sum of the two similarities for each pair of certificates
in the intersection of both match sets, and to then only classify
certificate pairs as final matches $\mathbf{M}_F$ that have an overall
similarity above a minimum similarity threshold $s_t$:
\begin{eqnarray}
\mathbf{M}_F & = & \{(C_i,C_k) \in \mathbf{M}_R \cap \mathbf{M}_G: \\
  \nonumber
  ~ & ~ &  (w_R sim_R(C_i,C_k) + w_G sim_G(C_i,C_k)) \ge s_t \},
\end{eqnarray}
where we set $w_R+w_G=1$, and $sim_R(\cdot,\cdot)$ and
$sim_G(\cdot,\cdot)$ are the relational and group similarities
between certificates as detailed in Sects.~\ref{sec-rel-sim}
and~\ref{sec-group-sim}, respectively.

Note that the overall minimum similarity threshold $s_t$ can be
different from the similarity threshold $s_m$ used in the pair-wise
linkage of individuals described in Sect.~\ref{sec-pair-wise}. While
$s_m$ needs to be set to a lower value to ensure as many matches as
possible are included into the set $\mathbf{S}$ of linked individuals
(i.e.\ high recall), $s_t$ can be set higher to ensure the final set
of matched certificates $\mathbf{M}_F$ has a high precision as well.

In the following section, using real historical data from Scotland, we
will experimentally evaluate how the different steps of our approach,
and the different ways of how to calculate relational and group
similarities, influence the quality of the final set $\mathbf{M}_F$ of
matched certificates.


\begin{table*}[t!]
  \centering
  \caption{Key characteristics of the Isle of Skye data sets used in
           the experiments. \label{table-data}}
  \begin{footnotesize}
  \begin{tabular}{|l|c|c|c|c|c|c|} \hline
  Data set & ~~Number of~~ & \multicolumn{5}{c|}{Number of unique
    values in selected attributes} \\ \cline{3-7}
  ~         & records / individuals & First name & Last name &
    Relationship  & Address & Occupation \\ \hline\hline
  Birth      & 17,614 / 70,456& 2,055 & 547 & 150 & 1,257 & ~~820 \\
  Marriage   & ~~2,668 / 16,008 & ~~427 & 343 & n/a & ~~903 &
    ~~589 \\
  Death      & 12,285 / 61,425 & ~~837 & 639 & 243 & ~~988 & ~~983 \\
  Census 1861 & 19,605 / 19,605 & ~~653 & 507 & 222 & 1,908 &
    1,318 \\ 
  Census 1871 & 18,102 / 18,102 & ~~611 & 459 & 146 & 1,924 & 1,559 \\
  Census 1881 & 17,684 / 17,684 &  ~~748 & 441 & 114 & 1,232 & 
    1,309 \\
  Census 1891 & 16,476 / 16,476 & ~~884 & 507 & 135 & 1,556 & 1,483 \\
  Census 1901~~ & 14,609 / 14,609 & ~~932 & 440 & 191 & 1,405 & 1,354
    \\ \hline
  \end{tabular}
  \end{footnotesize}
\end{table*}

\section{Experiments and Discussion}
\label{sec-data}

We evaluate our proposed approach using a collection of real Scottish
data sets that cover the population of the Isle of Skye over the
period from 1861 to 1901. These data sets have been extensively
curated and linked semi-manually (using database queries to extract
tables of possible links) by demographers who are experts in the
domain of linking such historical data~\cite{New11,Rei02}.

The linkage approach taken by the demographers followed long
established rules for family reconstruction~\cite{Wri73}, leading to
set of linked certificates that are heavily biased towards certain
types of links (as we discuss further below). As an example, a first
step in family reconstruction is often to link babies who had died
very shortly after birth (as these do not exhibit much variation or
changes in the names and addresses of their parents), followed by the
linking of birth certificates to their next census certificate (less
than ten years in the future for our data sets).

We thus have a set of manually generated links that allow us to
compare the quality and coverage of our automatically identified links
to those identified by the domain experts. We like to emphasize that
\emph{we do not have a gold standard of all true links in these data
sets} (neither do the demographers~\cite{Rei02}). All results
presented here therefore only indicate the overlap between manual and
automatically generated links between certificates.

As Table~\ref{table-data} shows, these data sets have some particular
characteristics. In common with other historical (census) data
sets~\cite{Ant14a,Fu14b,Fu12}, they have a very small number of unique
name values. The frequency distributions of names are also very
skewed. Across all data sets, the five most common first and last name
values occur in between $30\%$ and $40\%$ of all records. Many records
have missing values in address or occupation attributes, and
\emph{Informants} are mostly missing (we therefore do not include any
records with this role).

While we do have a set of manually generated links available (biased
as described above), we only use them for evaluation purpose because
our overall aim is to develop fully unsupervised techniques that are
applicable even when no training data are available, as is often the
case in practical record linkage projects~\cite{Chr12}. The set of
manually linked certificates contains $54,537$ life segments, where
$19,377$ consist of a single certificate and the maximum number of
certificates in a life segment is $8$, leading to $84,895$ manually
linked certificate pairs. The median number of certificates per life
segment is $2$ and the average is $2.41$.

\begin{table}[t!]
  \centering
  \caption{Distribution of manually linked certificate pairs of
           different types. \label{table-link-types}}
  \begin{footnotesize}
  \begin{tabular}{|l|r|r|} \hline
   Certificate pair types & Number of links & Percentage \\
    \hline\hline
    All               & 84,895~~~~~~ &  100\%~~~ \\
    Birth--Census     & 24,045~~~~~~ & 28.3\%~~~ \\
    Birth--Death      &  3,254~~~~~~ &  3.8\%~~~ \\
    Birth--Marriage   &    438~~~~~~ &  0.5\%~~~ \\
    Census--Census    & 30,372~~~~~~ & 35.8\%~~~ \\
    Census--Death     & 15,385~~~~~~ & 18.1\%~~~ \\
    Census--Marriage  & 10,550~~~~~~ & 12.4\%~~~ \\
    Death--Marriage   &    747~~~~~~ &  0.9\%~~~ \\
    Marriage-Marriage &    104~~~~~~ &  0.1\%~~~ \\ \hline
  \end{tabular}
  \end{footnotesize}
\end{table}

As Table~\ref{table-link-types} shows, the number of different types
of manually linked certificate pairs is highly skewed, with the
majority of true matches being between pairs of census (35.8\%), birth
to census (28.3\%), and census to death (18.1\%) certificates. There
are less than a few hundred links between birth and marriage and death
and marriage certificates, and an even smaller number of marriage to
marriage links. Understandably, there are no true matching pairs of
certificates of types birth to birth and death to death in the set of
manually linked certificates.



To compare the quality of our automatic linkage approach with the
manually identified links, we use \emph{precision}, $p = tp / (tp+fp)$
and \emph{recall}, $r = tp/(tp+fn)$ as commonly used in record linkage
evaluations~\cite{Chr12}. 
We set as \emph{true positives}, $tp$, the set of certificate pairs
linked both manually and automatically; as \emph{false positives},
$fp$, those pairs linked only automatically by our approach but not
manually by the domain experts; and as \emph{false negatives}, $fn$,
those pairs linked manually by the domain experts but not by our
approach.

\begin{figure*}[t!]
  \centering
  \includegraphics[width=0.49\textwidth]
                  {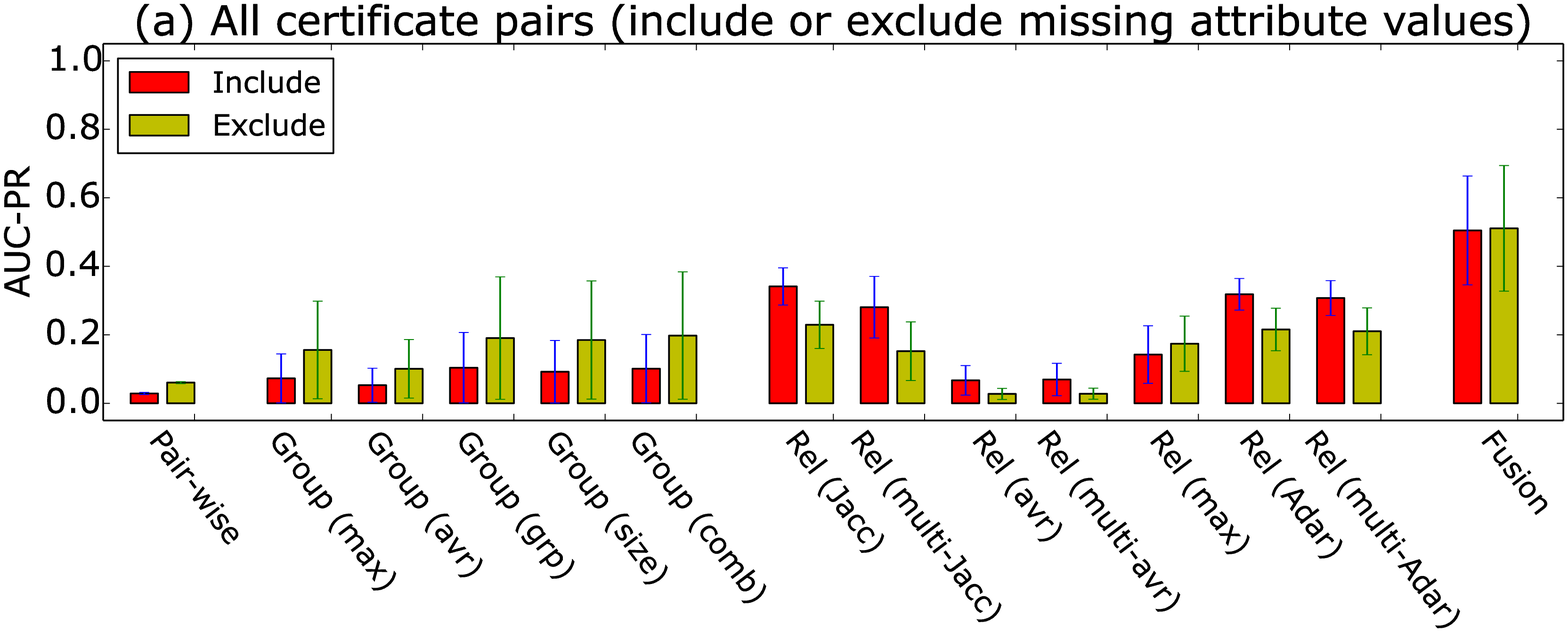}~~~~
  \includegraphics[width=0.49\textwidth]
                  {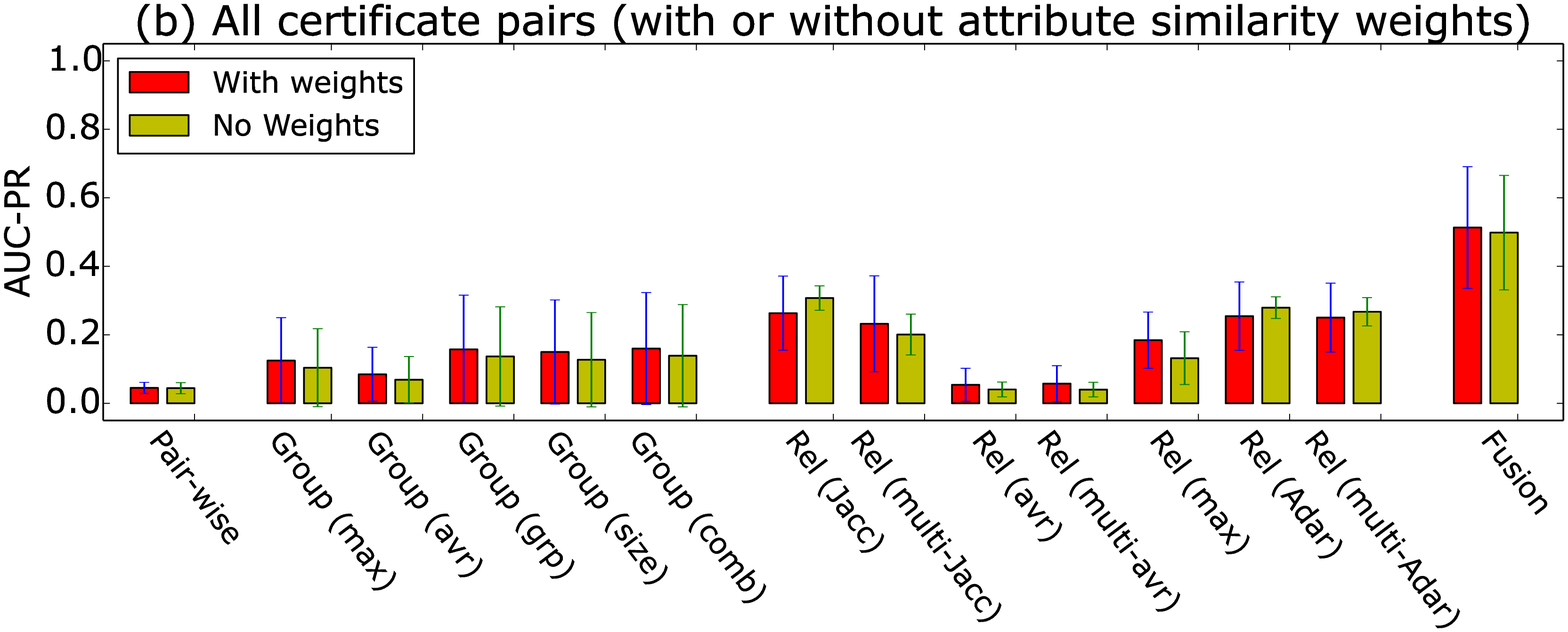}
  ~ \\[-2mm]
  \includegraphics[width=0.49\textwidth]
                  {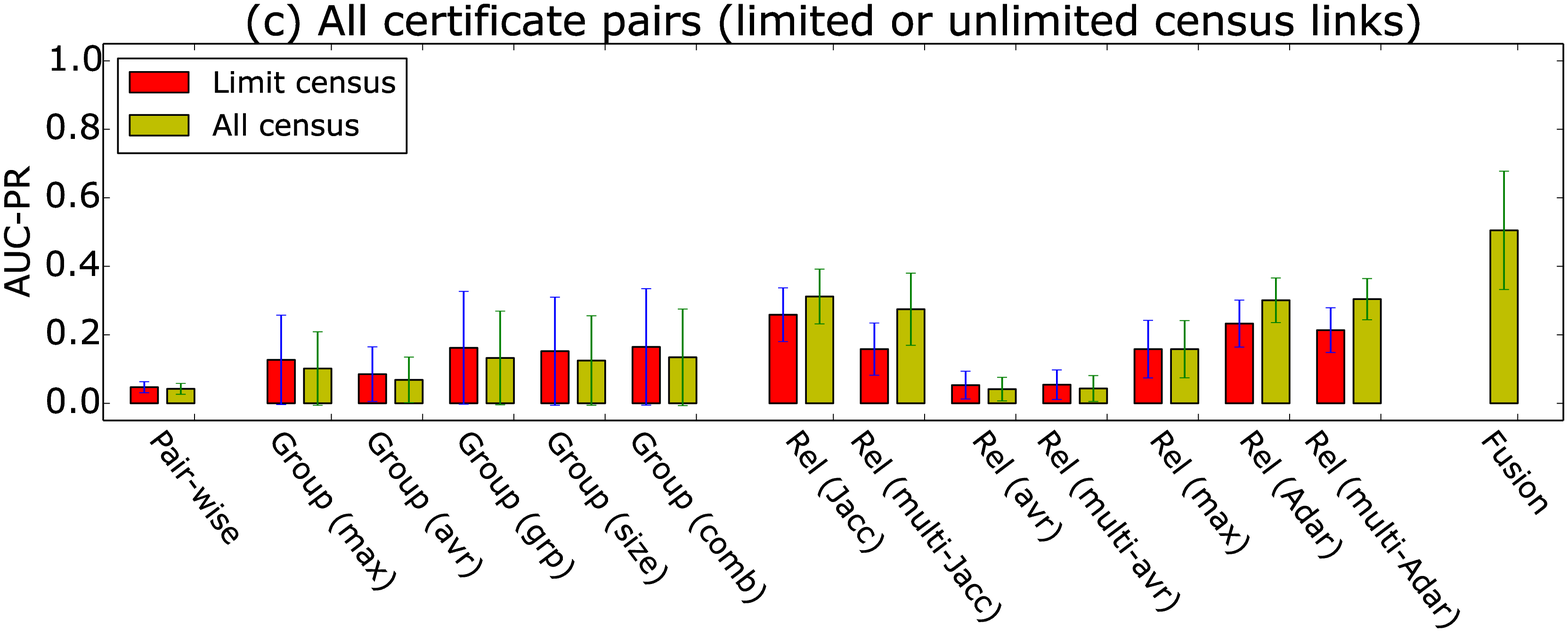}~~~~
  \includegraphics[width=0.49\textwidth]
                  {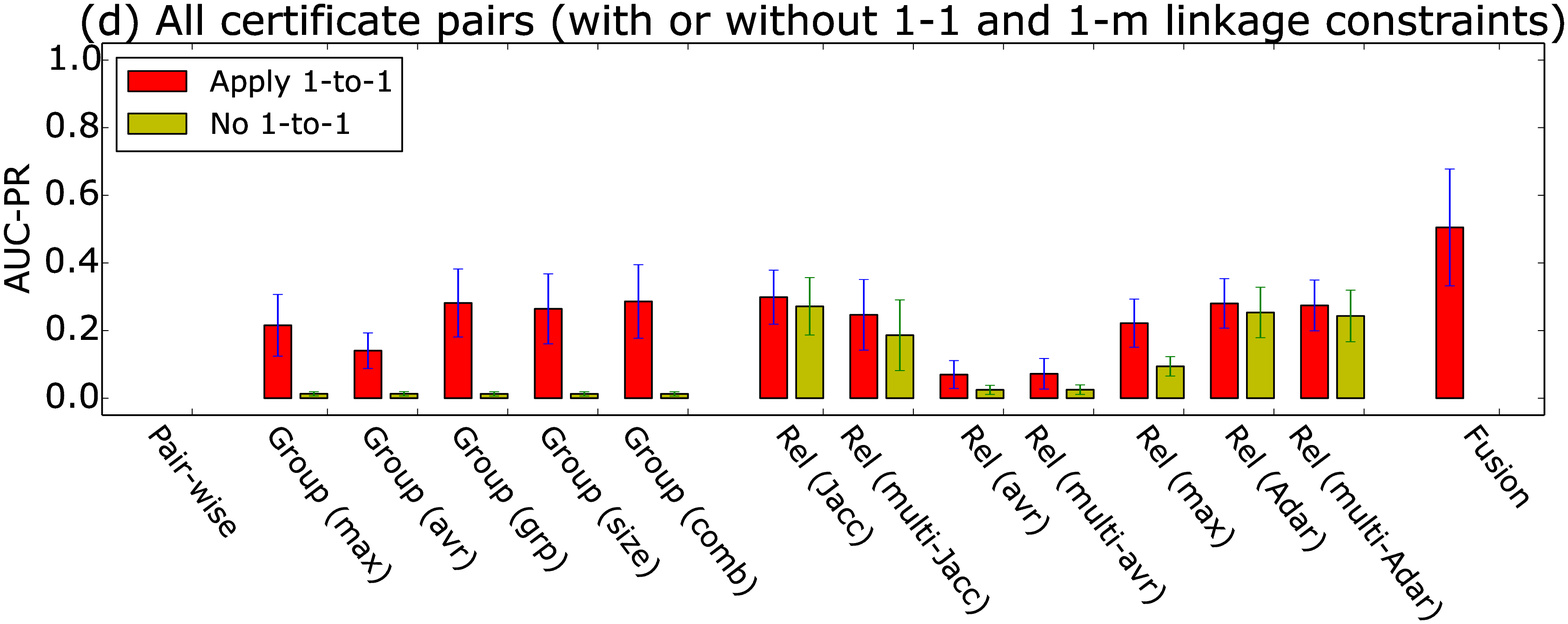}
  \caption{Area under the precision-recall curve (AUC-PR) results
           for all types of certificate pairs and different linkage
           options as discussed in Sect.~\ref{sec-data}.
           \label{fig-res-all}}
\end{figure*}

Record linkage is often a very imbalanced problem~\cite{Chr12}. In our
case the ratio of the number of manually linked certificate pairs over
the number of compared certificates
is $1:326$. This makes identifying linked certificates highly
challenging. To provide summary information for our automatic versus
manual linkage approaches, we report the area under the
precision-recall curve (AUC-PR) which is more meaningful for such an
imbalanced problem compared to the area under the ROC
curve~\cite{Dav06}. We use the Scikit-learn
(\url{http://scikit-learn.org})
machine learning package for calculating AUC-PR~\cite{Ped11}.

We implemented our approach in Python 2.7 using the functionalities of
the \emph{Febrl}~\cite{Chr09b} record linkage system. We conducted all
experiments on a server with 64-bit Intel Xeon 2.4 GHz CPUs, 128
GBytes of memory, and running Ubuntu 14.04. To facilitate
repeatability we will make our programs available, while for the Isle
of Skye data sets the interested reader is encouraged to contact the
owners~\cite{Rei02}.

We applied basic cleaning of attribute values in our data by removing
unwanted characters, standardizing age and gender values, and imputing
birth years calculated from age values. In the blocking step we used
$|B|= 3,950$ blocking keys (different combinations of role types,
selected attributes, and phonetic encodings). We set the minimum
individual similarity threshold as $s_m=0.4$, resulting in a total of
$|\mathbf{S}| = 27,689,015$ compared pairs of individual records in
the attribute similarity calculation step (step 3).

\newpage

We also investigated the following four options of how to calculate
pair-wise similarities and constraint comparisons:
\begin{itemize}
\item Given the large number of missing values in certain attributes
  (notably in addresses and occupations), when calculating individual
  similarities we either included (with a similarity of $0$) or
  excluded attributes when at least one value in an attribute pair was
  missing.
\item Also in the pair-wise similarity step, we either assigned each
  attribute a weight of $1.0$ towards its contribution to the
  individual record pair similarity $sim(r_{i,j},r_{k,l})$ (as described
  in Sect.~\ref{sec-pair-wise}), or we calculated attribute specific
  weights as is commonly done in traditional record
  linkage~\cite{Her07} using the manual linked certificates (with
  higher weights given to attributes that are more similar in manually
  linked certificate pairs compared to randomly selected pairs that do
  not correspond to the same individuals).
\item We either allowed links to census certificates without temporal
  constraints, or we limited such links to only one decade
  before/after a census, with the aim of limiting links to only the
  temporal closest census certificates. For example, a birth
  certificate from 1868 would only be linked to census certificates in
  1861 and 1871. The hypothesis is that such links should be of higher
  quality because households and families are changing over time and
  thereby become more difficult to link over longer time periods.
\item Finally, we investigated how the 1-to-1 and 1-to-many linkage
  constraints described in Sect.~\ref{sec-temp-1-1} (when enforced or
  not enforced) affect the quality of the obtained set of linked
  certificates.
\end{itemize}

\begin{figure*}[t!]
  \centering
  \includegraphics[width=0.49\textwidth]
                  {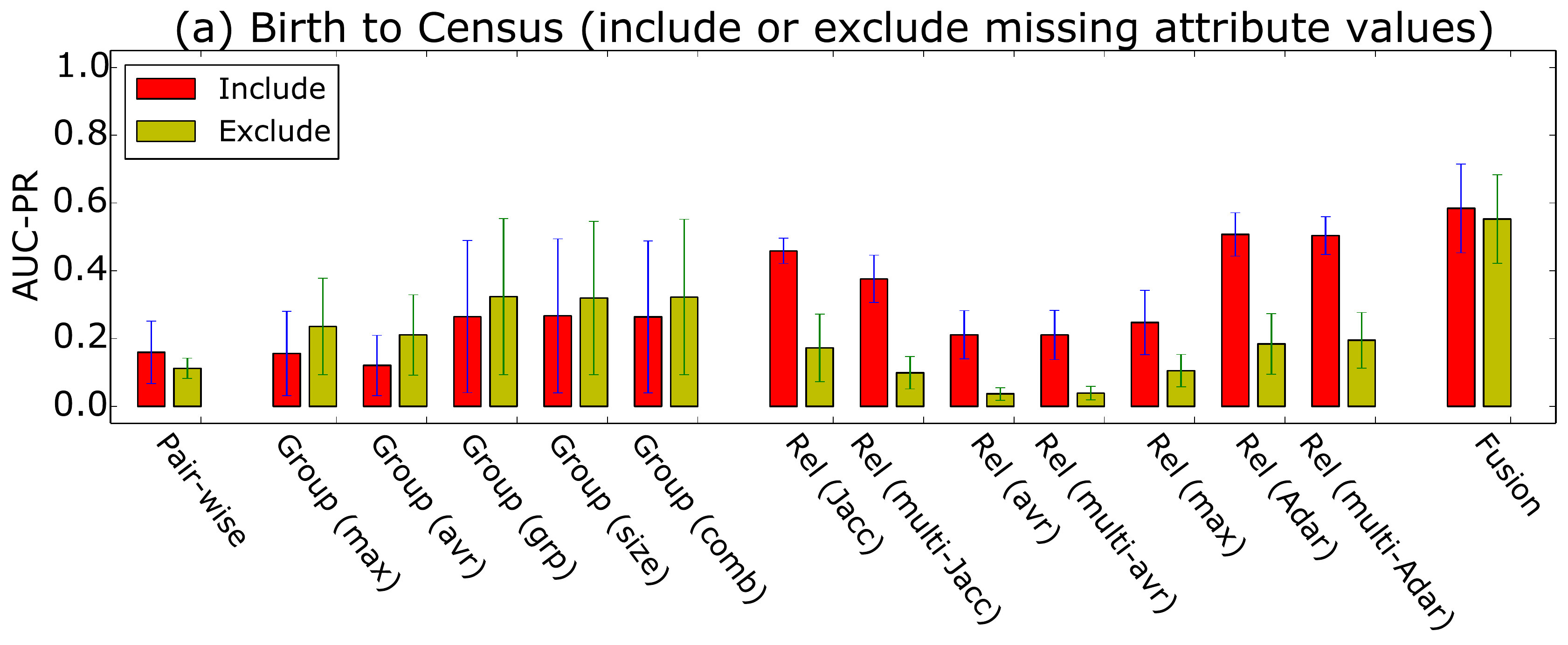}~~~~
  \includegraphics[width=0.49\textwidth]
                  {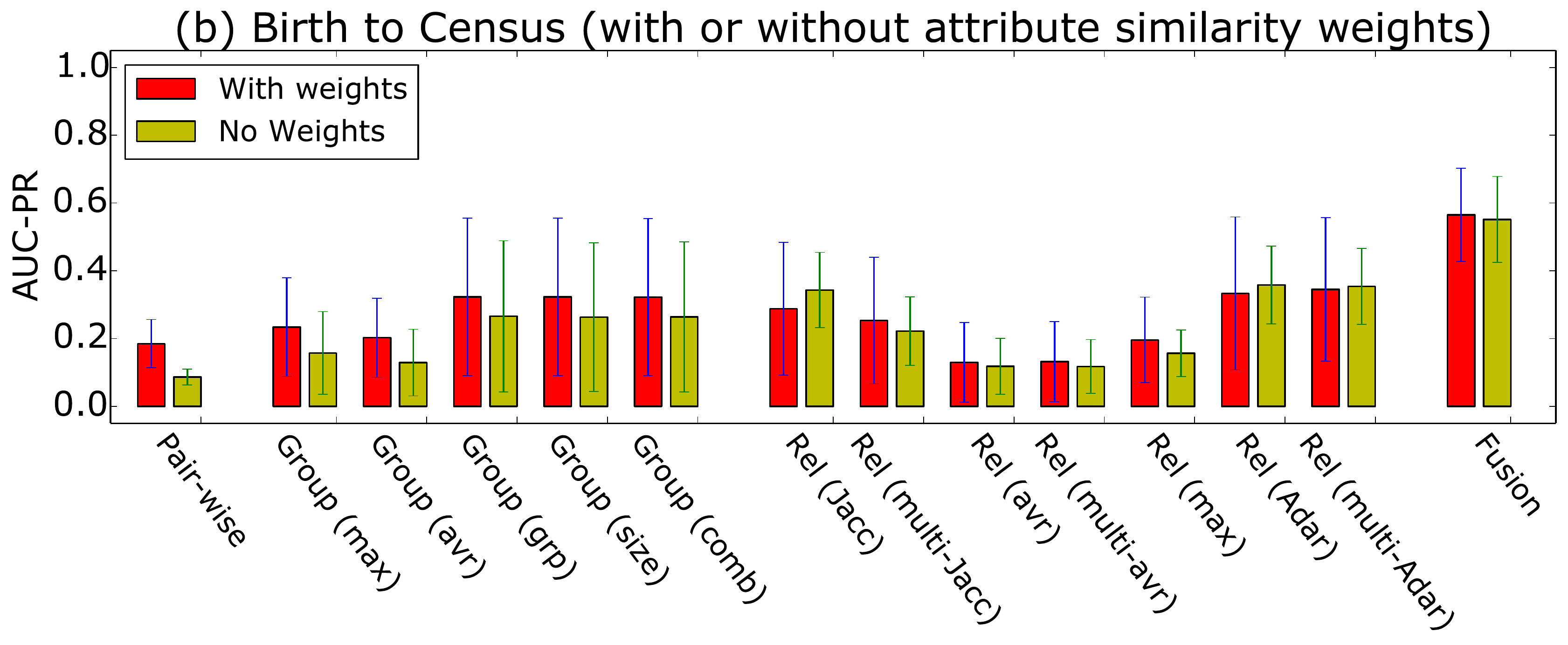}
  ~ \\[-2mm]
  \includegraphics[width=0.49\textwidth]
                  {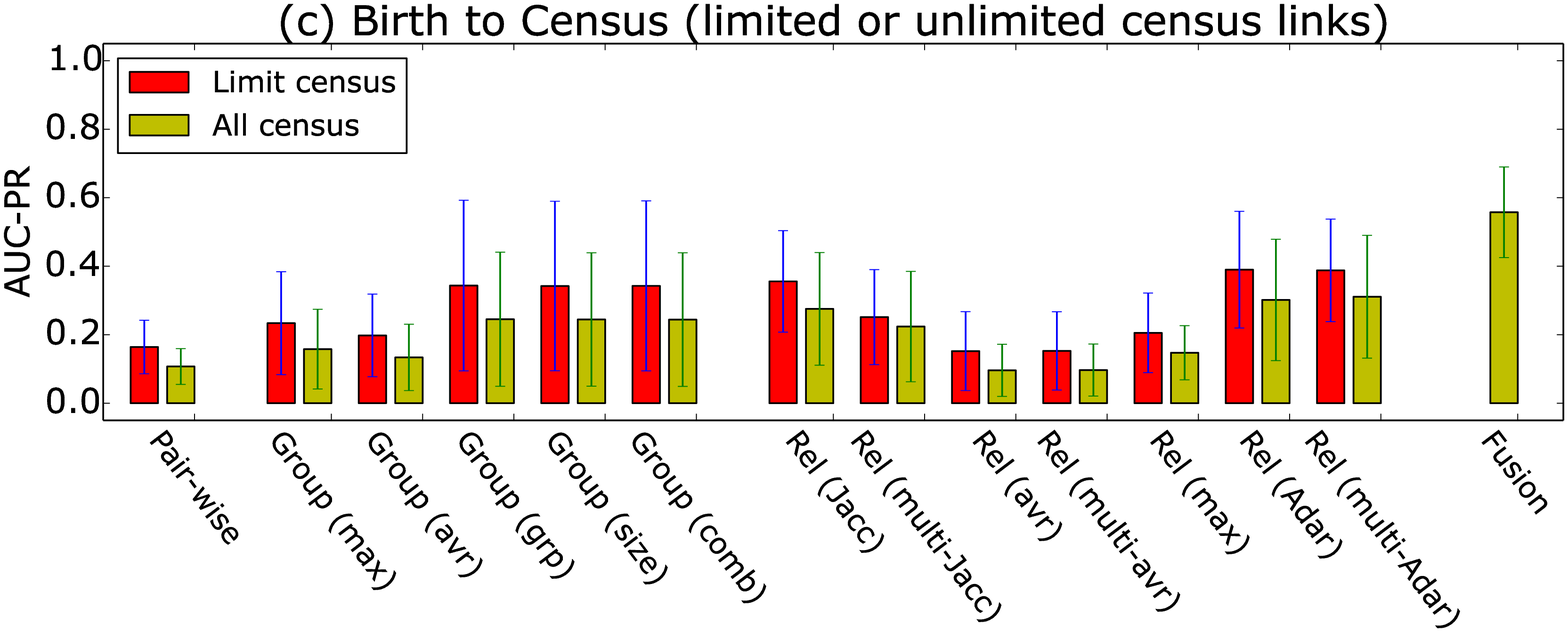}~~~~
  \includegraphics[width=0.49\textwidth]
                  {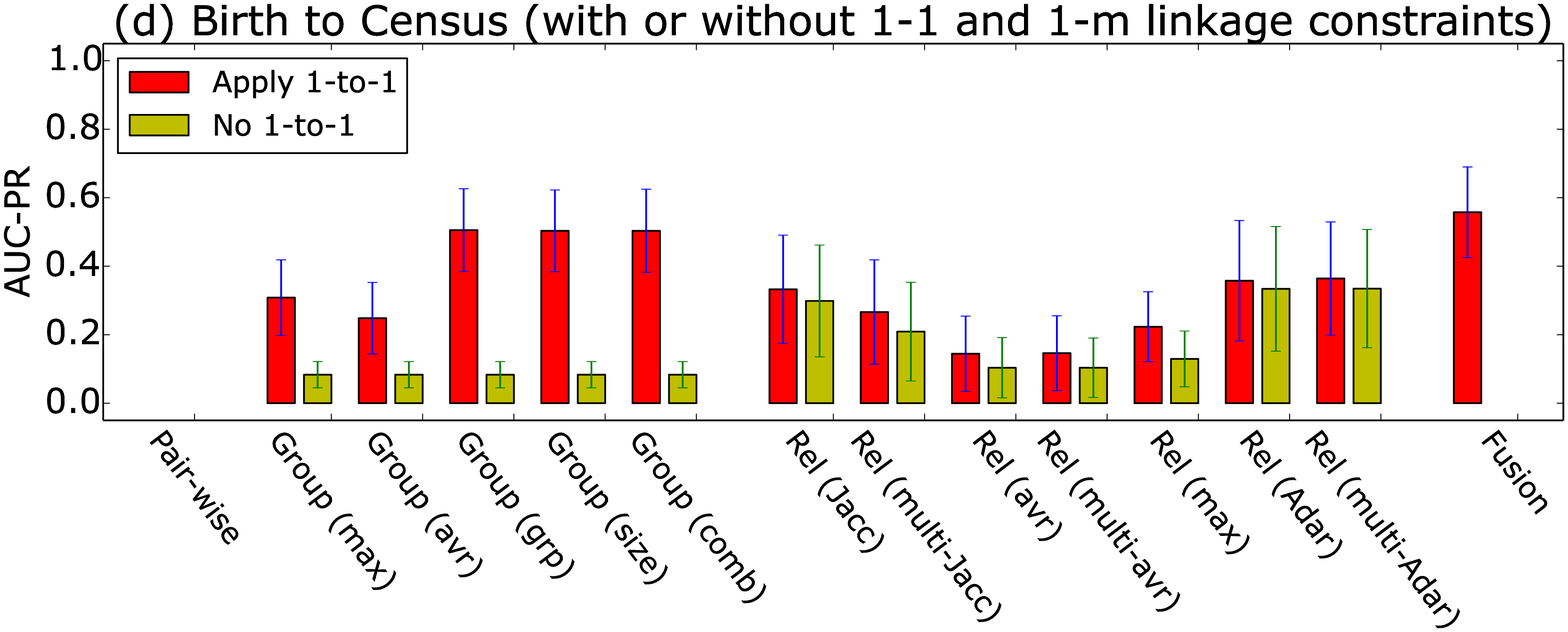}
  \caption{Area under the precision-recall curve (AUC-PR) results
           for birth to census certificate pairs and different linkage
           options as discussed in Sect.~\ref{sec-data}.
           \label{fig-res-b-c}}
\end{figure*}

\begin{figure*}[t!]
  \centering
  \includegraphics[width=0.49\textwidth]
                  {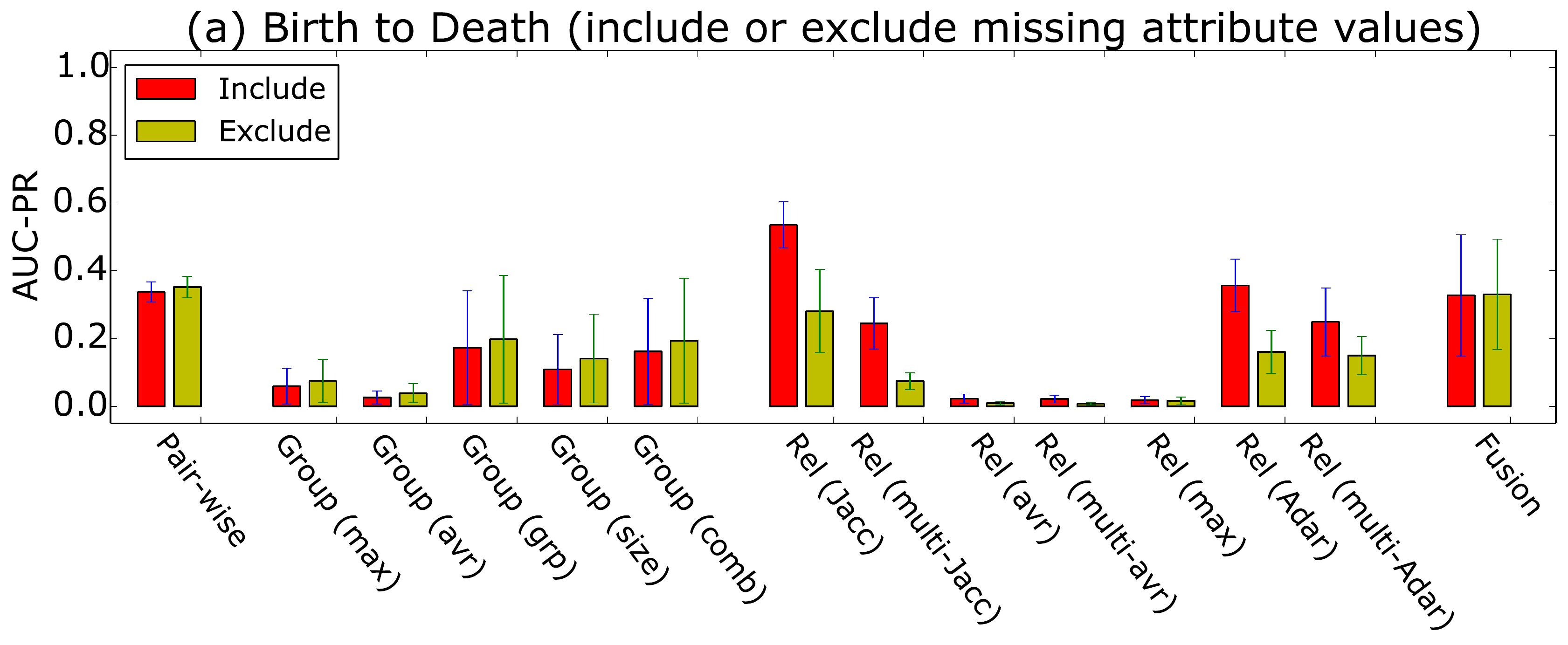}~~~
  \includegraphics[width=0.49\textwidth]
                  {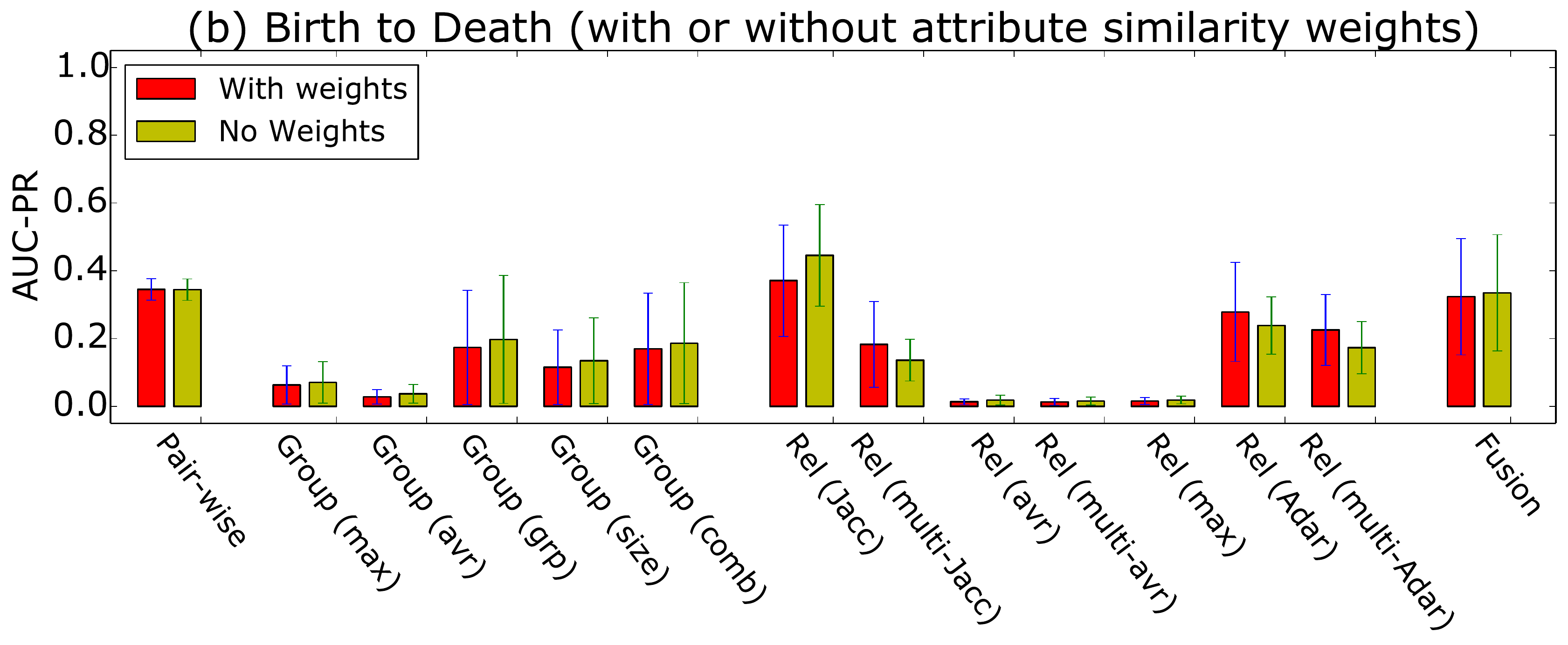}
  ~ \\[-2mm]
\includegraphics[width=0.49\textwidth]
                  {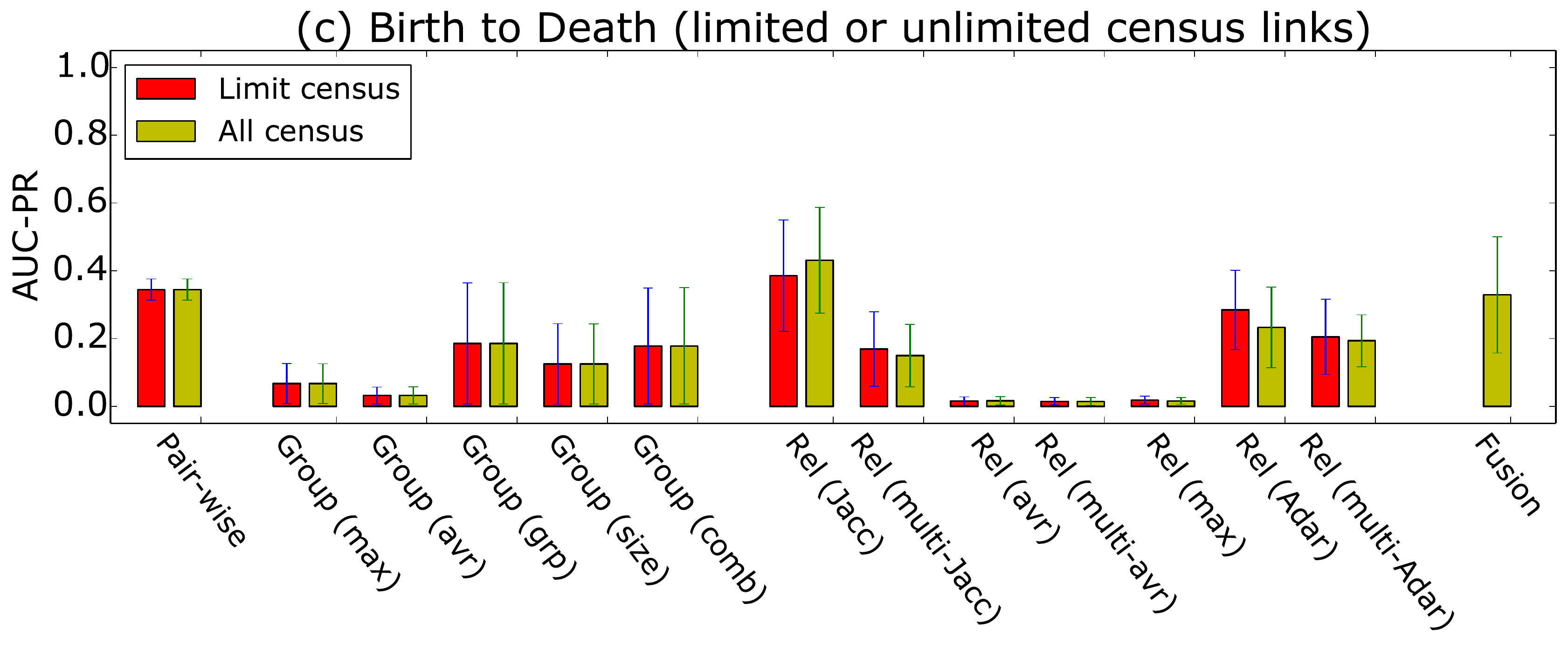}~~~
  \includegraphics[width=0.49\textwidth]
                  {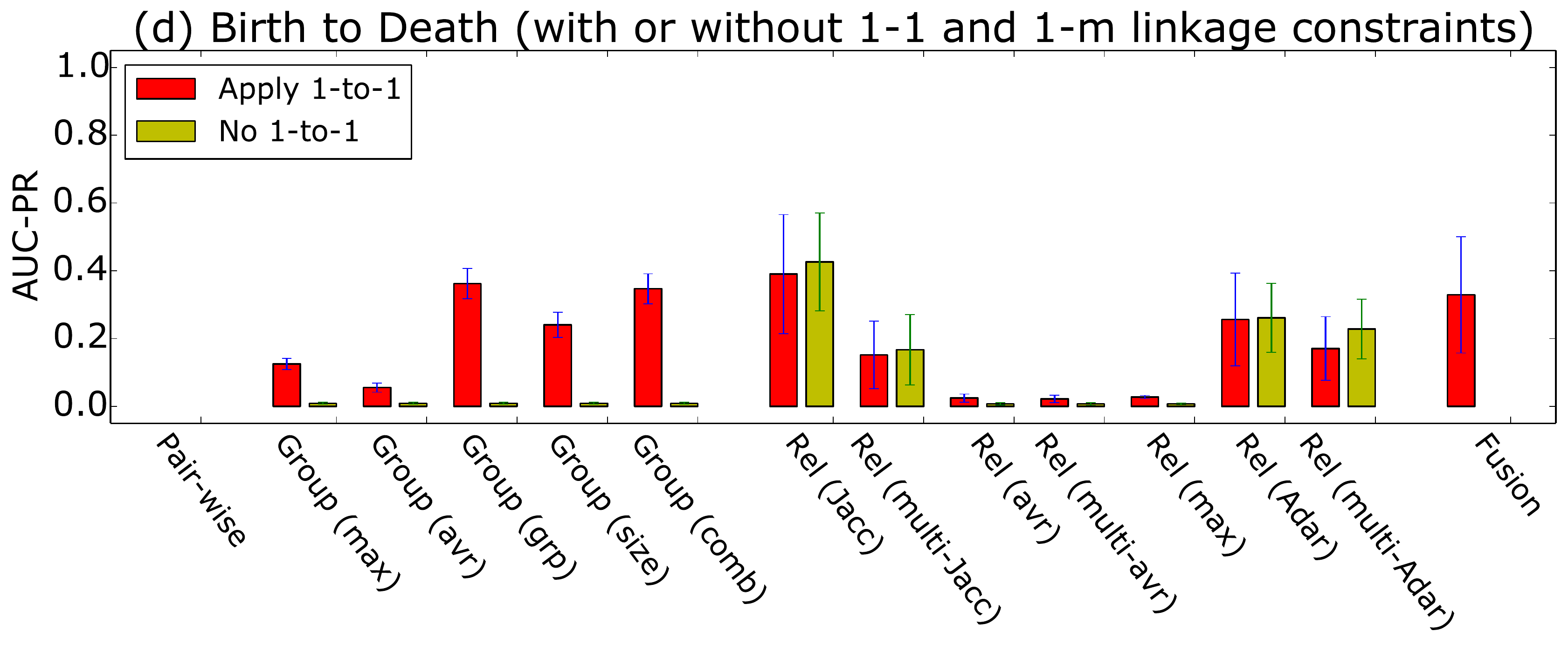}
  \caption{Area under the precision-recall curve (AUC-PR) results
           for birth to death certificate pairs and different linkage
           options as discussed in Sect.~\ref{sec-data}.
           \label{fig-res-b-d}}
\end{figure*}

\begin{figure*}[t!]
  \centering
  \includegraphics[width=0.49\textwidth]
                  {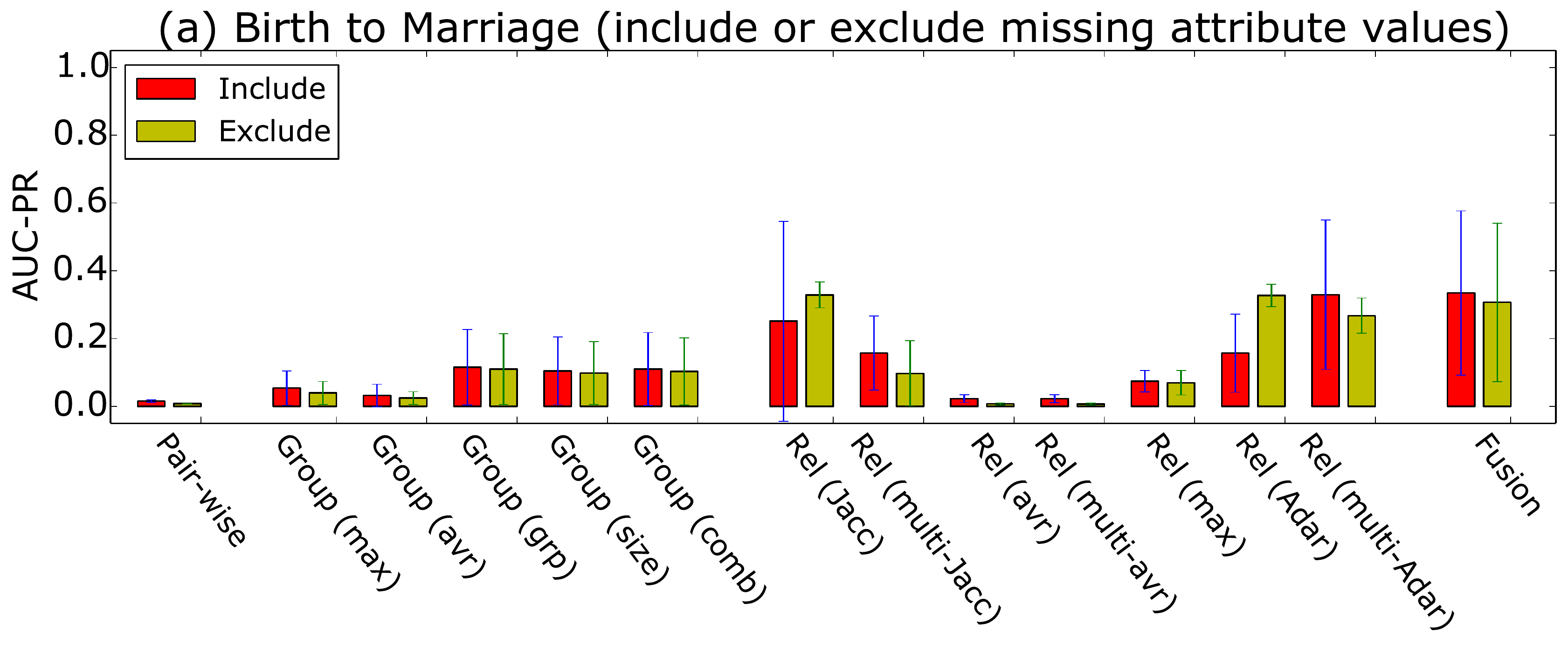}~~
  \includegraphics[width=0.49\textwidth]
                  {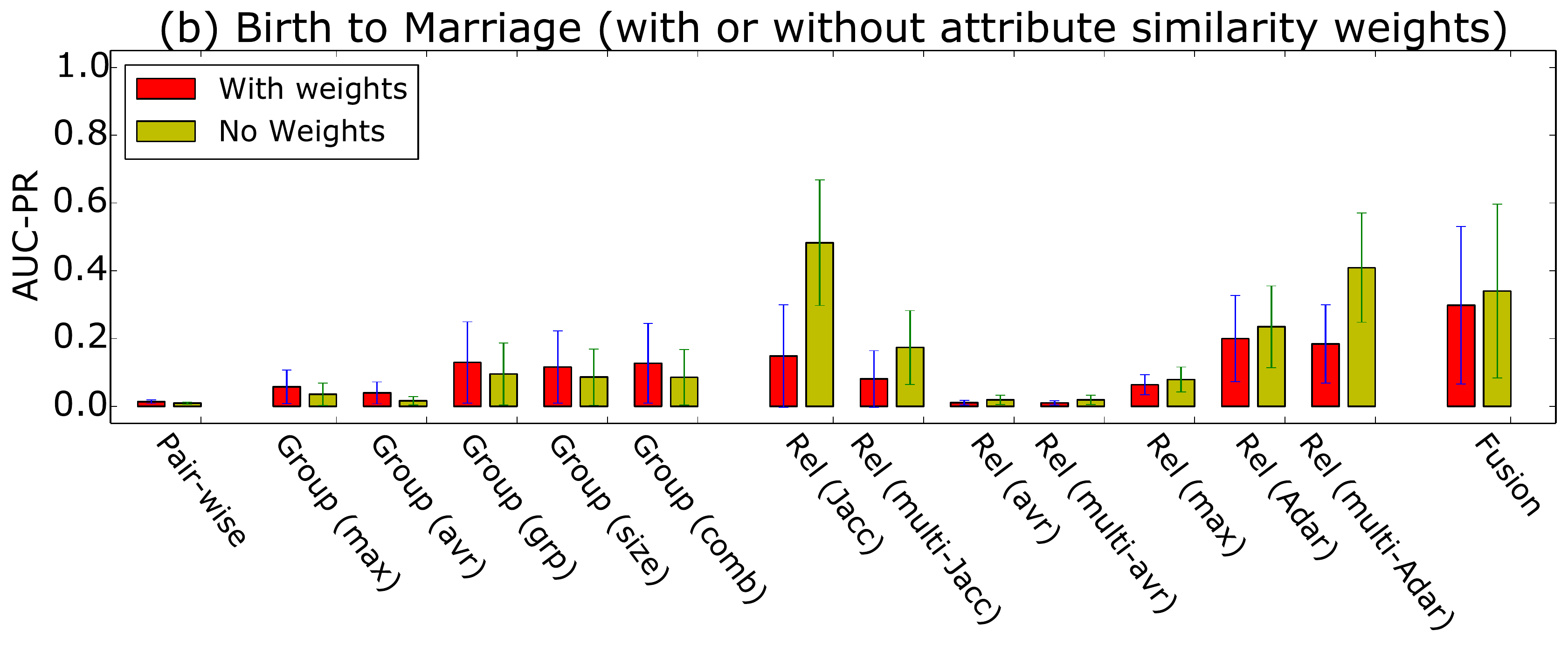}
  ~ \\[-2mm]
  \includegraphics[width=0.49\textwidth]
                  {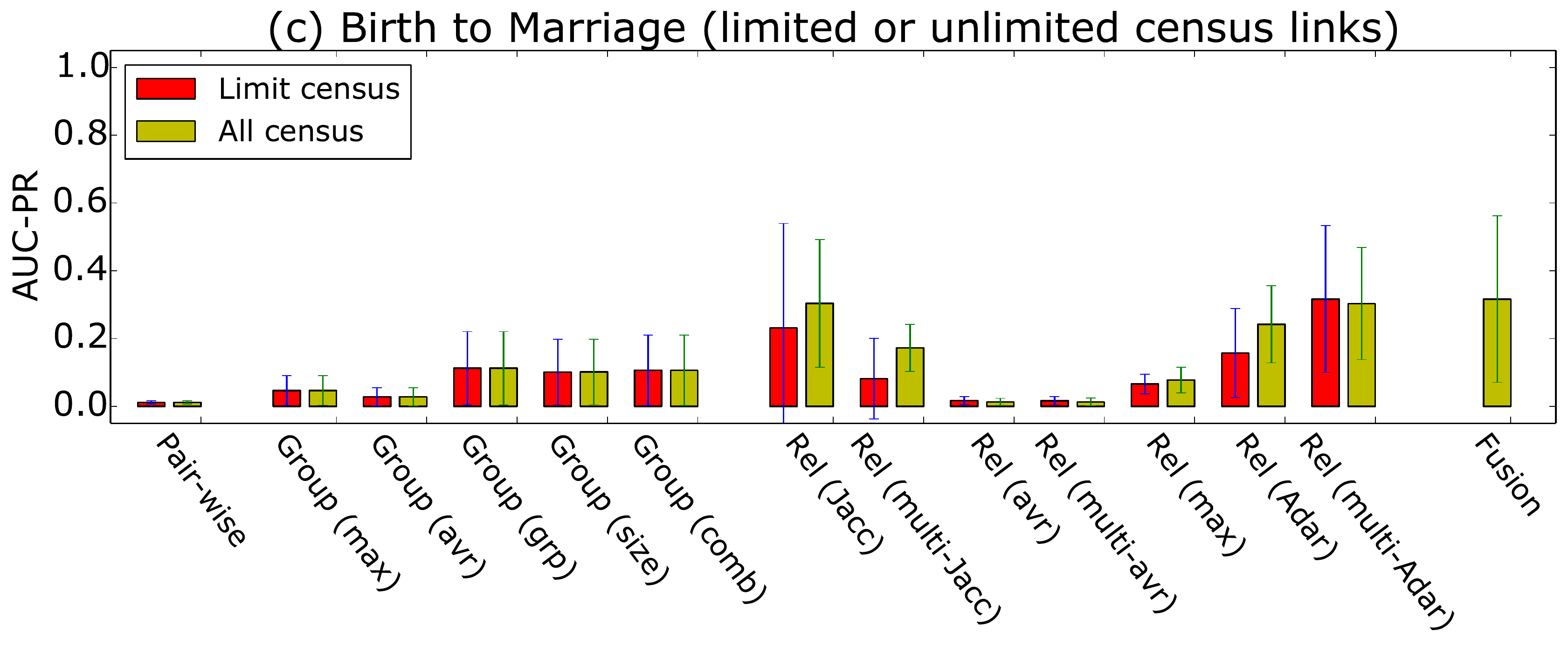}~~
  \includegraphics[width=0.49\textwidth]
                  {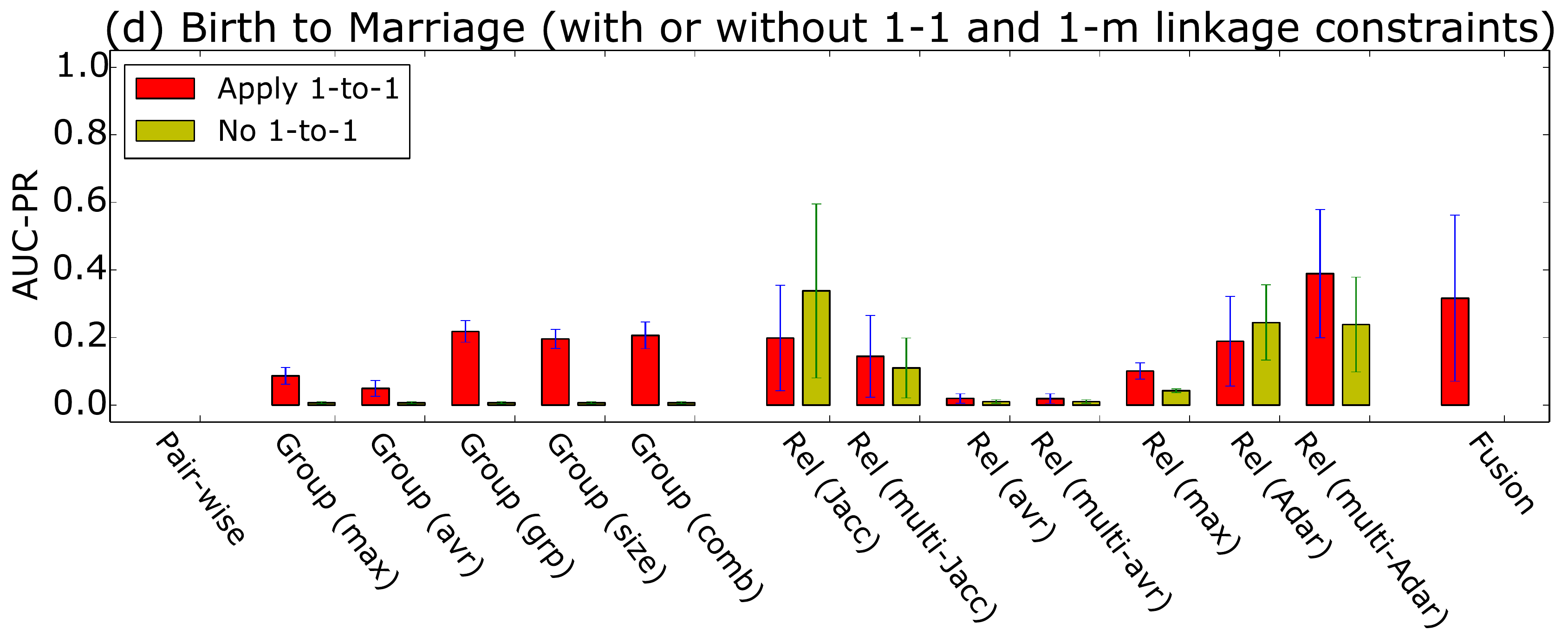}
  \caption{Area under the precision-recall curve (AUC-PR) results
           for birth to marriage certificate pairs and different
           linkage options as discussed in Sect.~\ref{sec-data}.
           \label{fig-res-b-m}}
\end{figure*}

\begin{figure*}[t!]
  \centering
  \includegraphics[width=0.49\textwidth]
                  {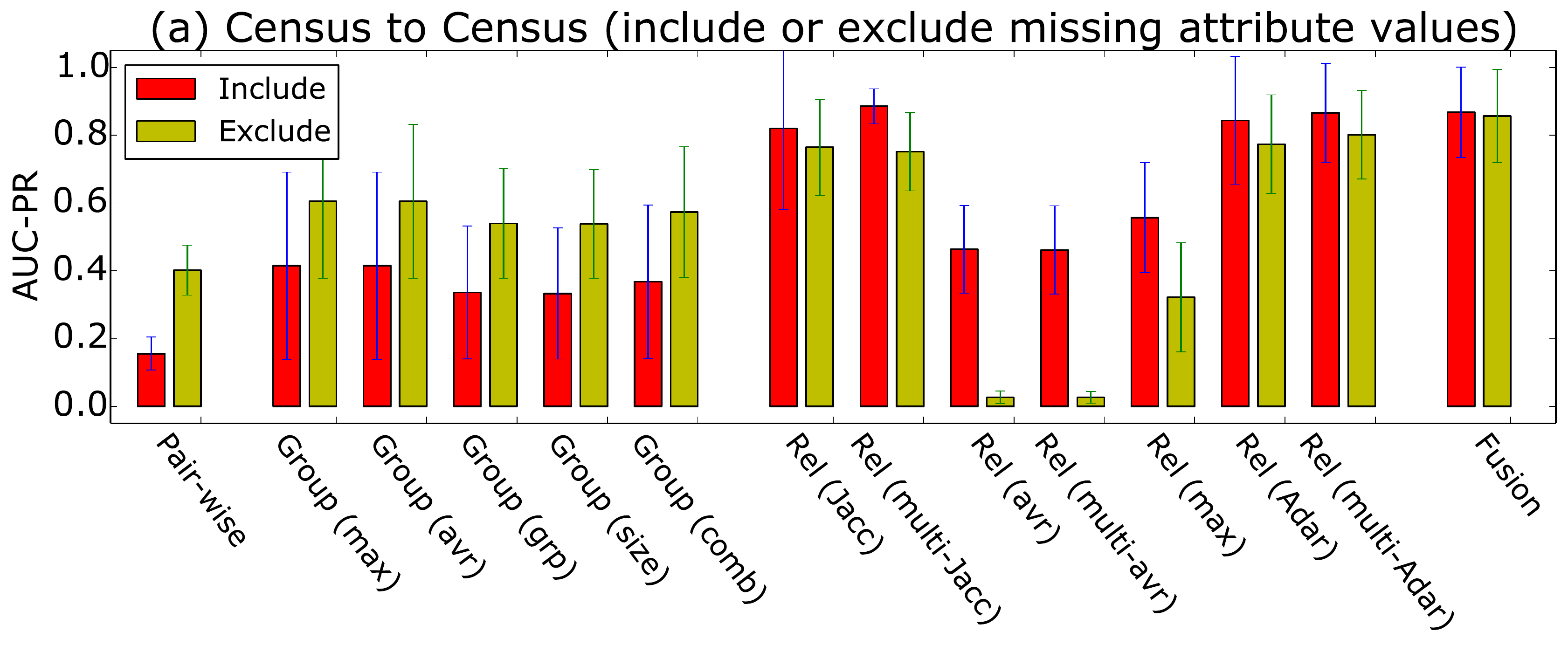}~~~~
  \includegraphics[width=0.49\textwidth]
                  {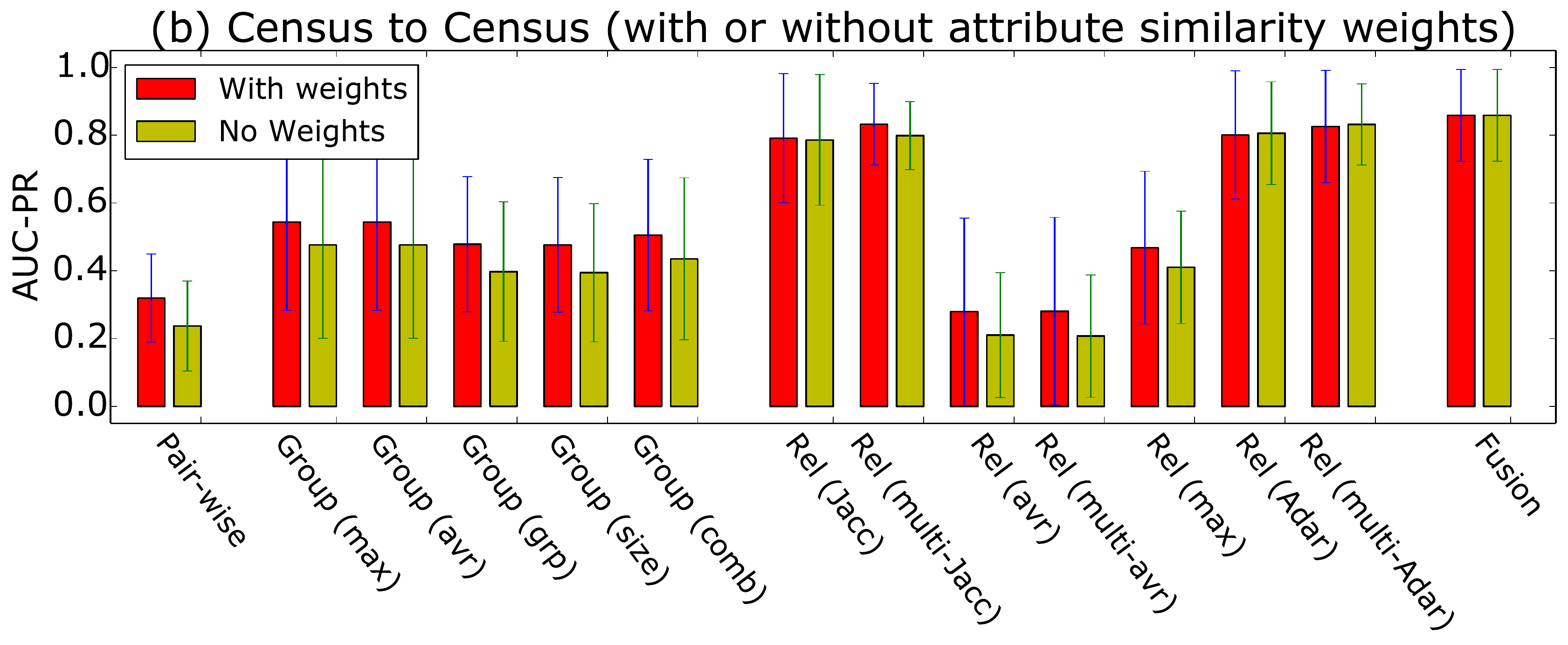}
  ~ \\[-2mm]
  \includegraphics[width=0.49\textwidth]
                  {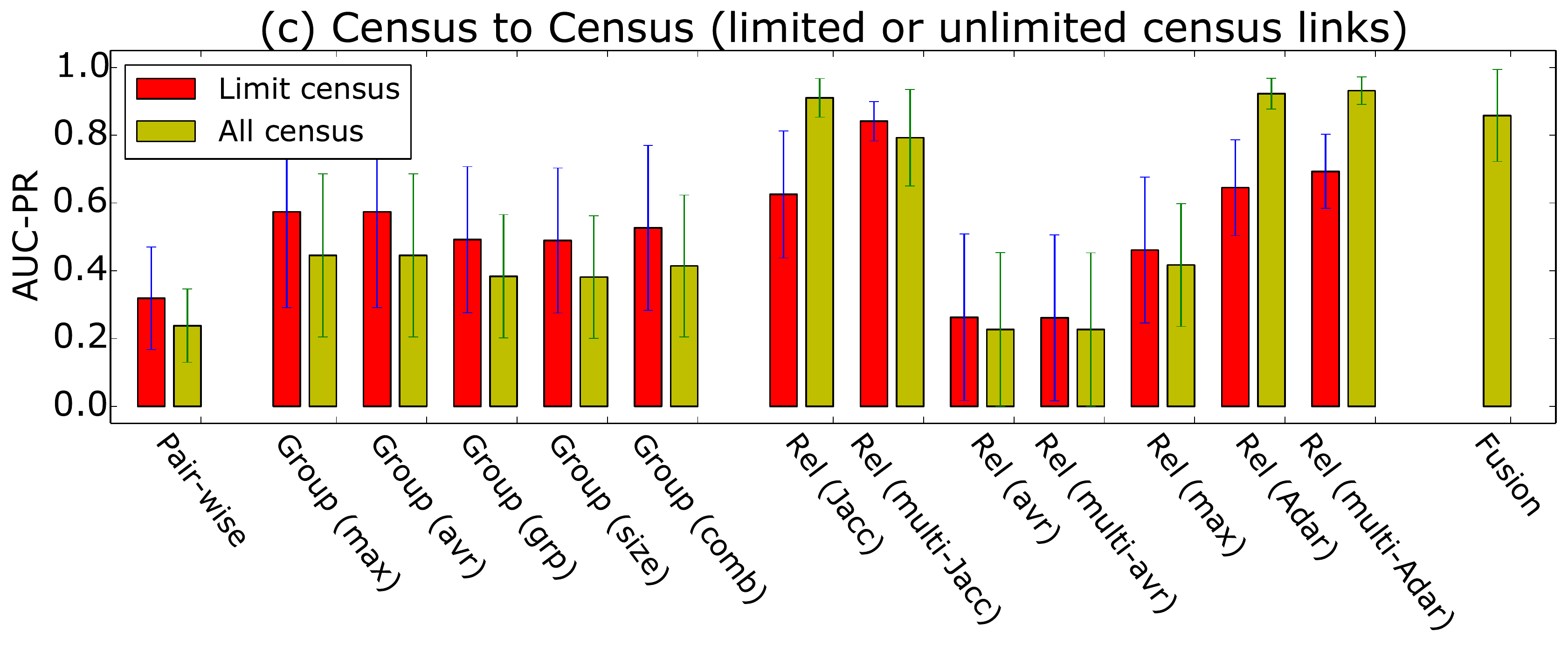}~~~~
  \includegraphics[width=0.49\textwidth]
                  {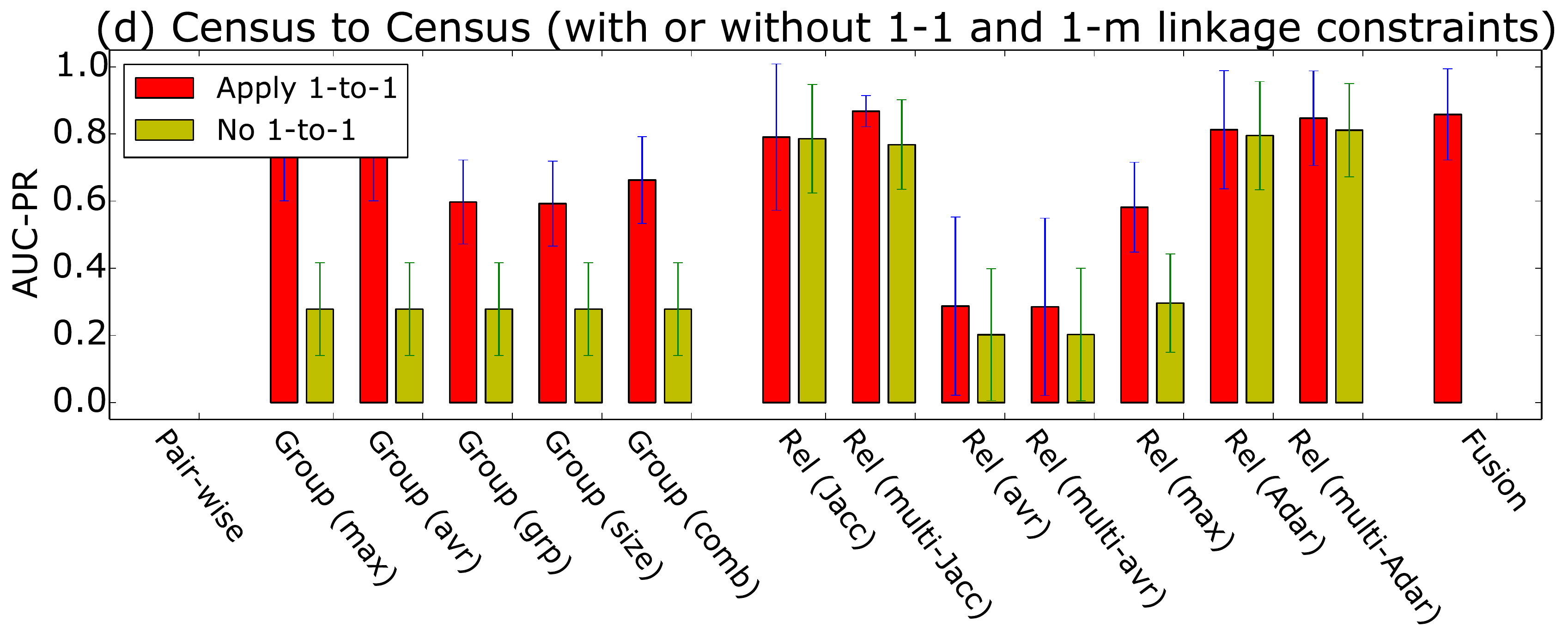}
  \caption{Area under the precision-recall curve (AUC-PR) results
           for census to census certificate pairs and different
           linkage options as discussed in Sect.~\ref{sec-data}.
           \label{fig-res-c-c}}
\end{figure*}

\begin{figure*}[t!]
  \centering
  \includegraphics[width=0.49\textwidth]
                  {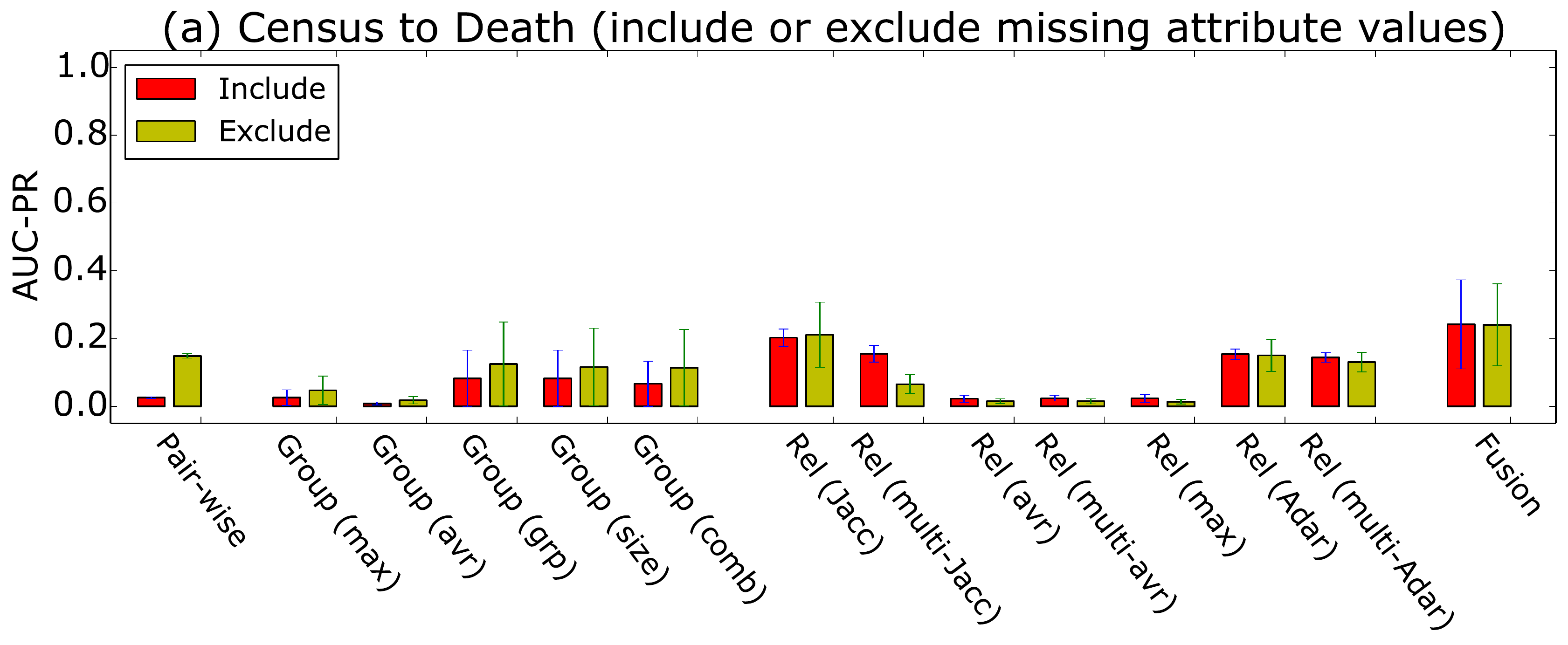}~~~~
  \includegraphics[width=0.49\textwidth]
                  {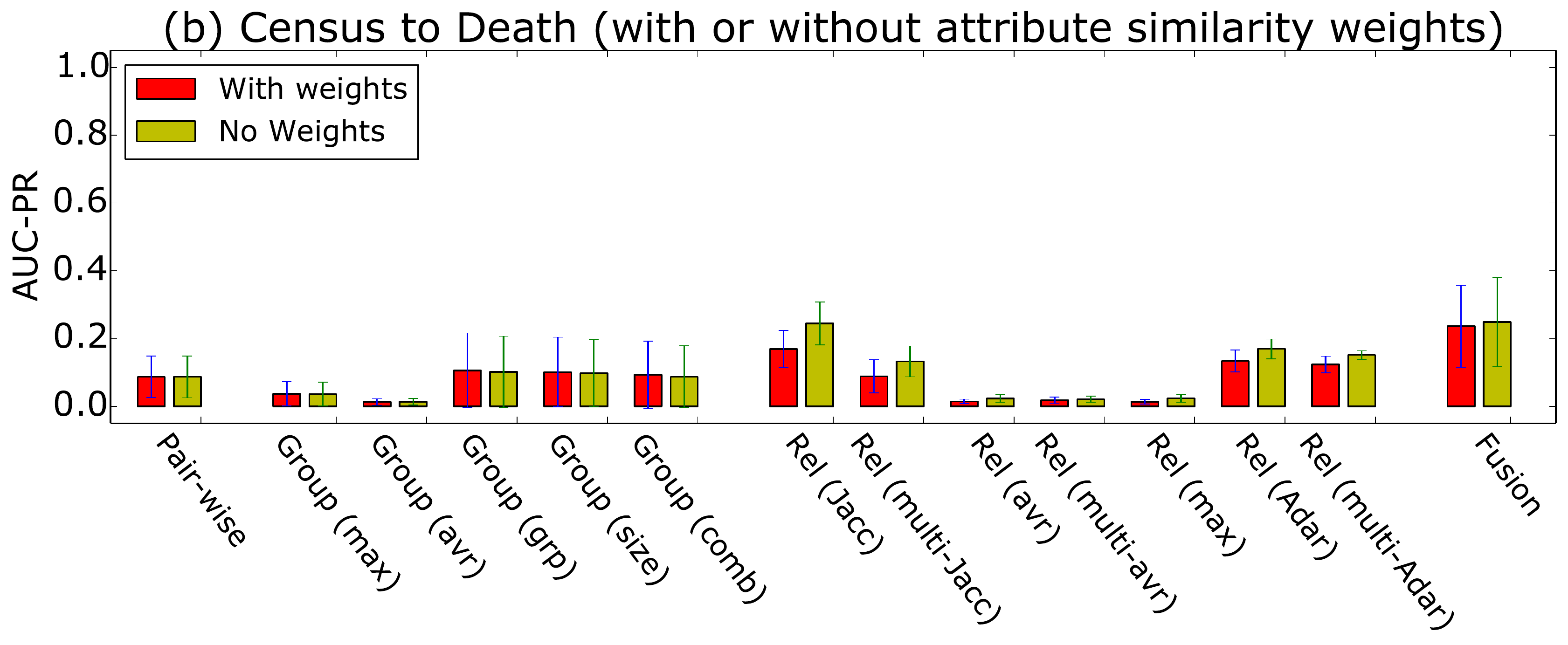}
  ~ \\[-2mm]
  \includegraphics[width=0.49\textwidth]
                  {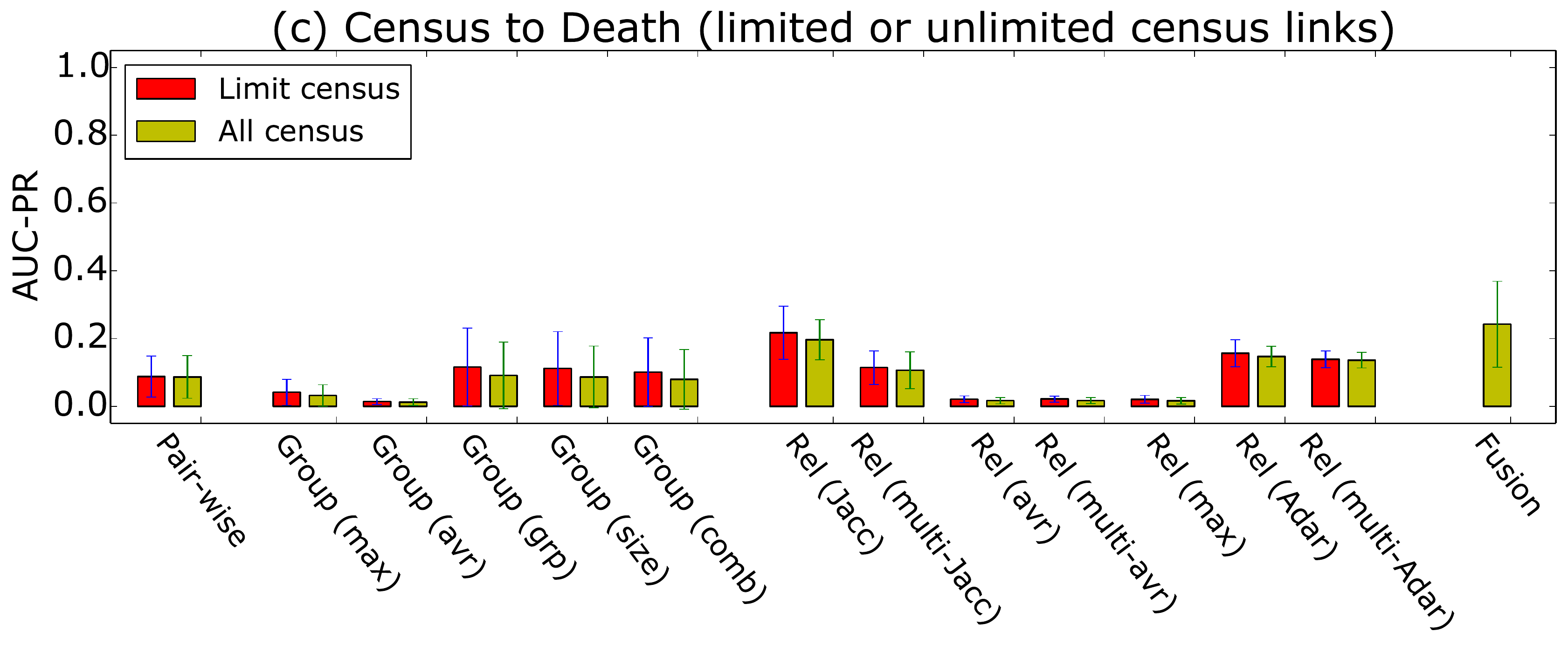}~~~~
  \includegraphics[width=0.49\textwidth]
                  {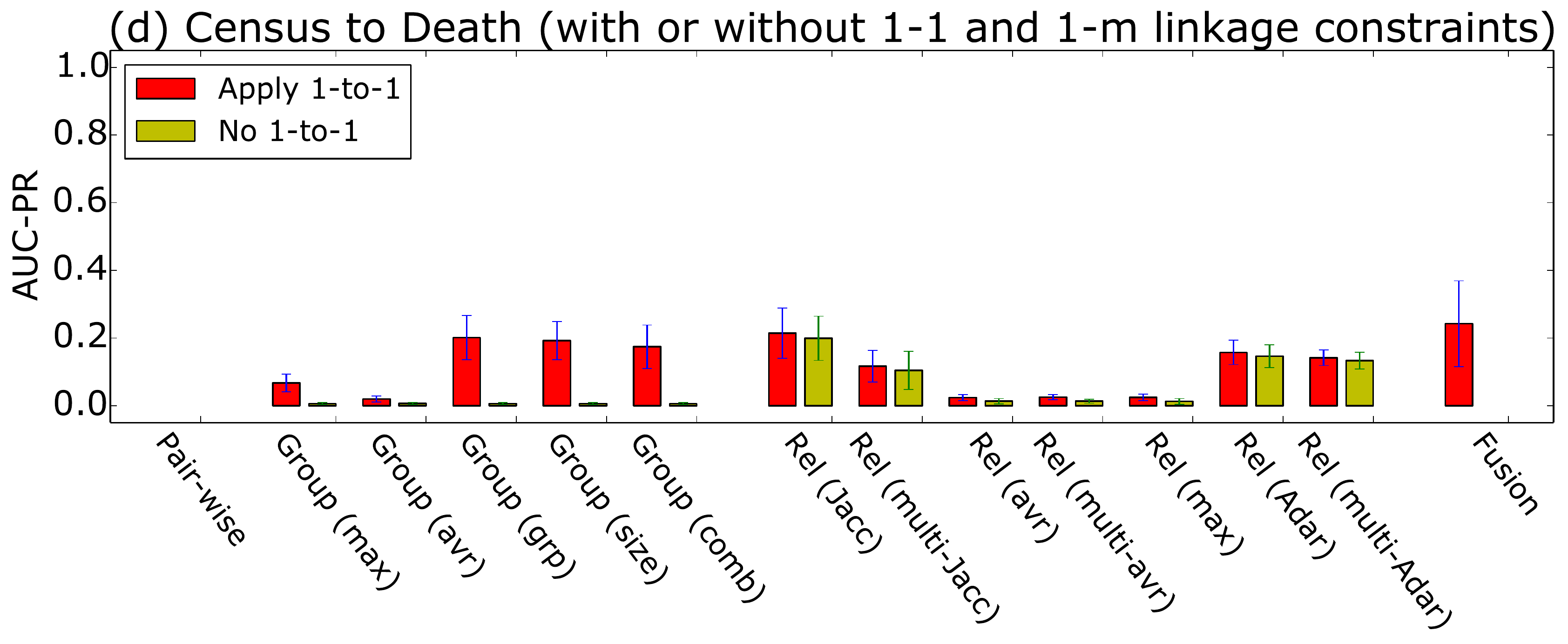}
  \caption{Area under the precision-recall curve (AUC-PR) results
           for census to death certificate pairs and different linkage
           options as discussed in Sect.~\ref{sec-data}.
           \label{fig-res-c-d}}
\end{figure*}

\begin{figure*}[t!]
  \centering
  \includegraphics[width=0.49\textwidth]
                  {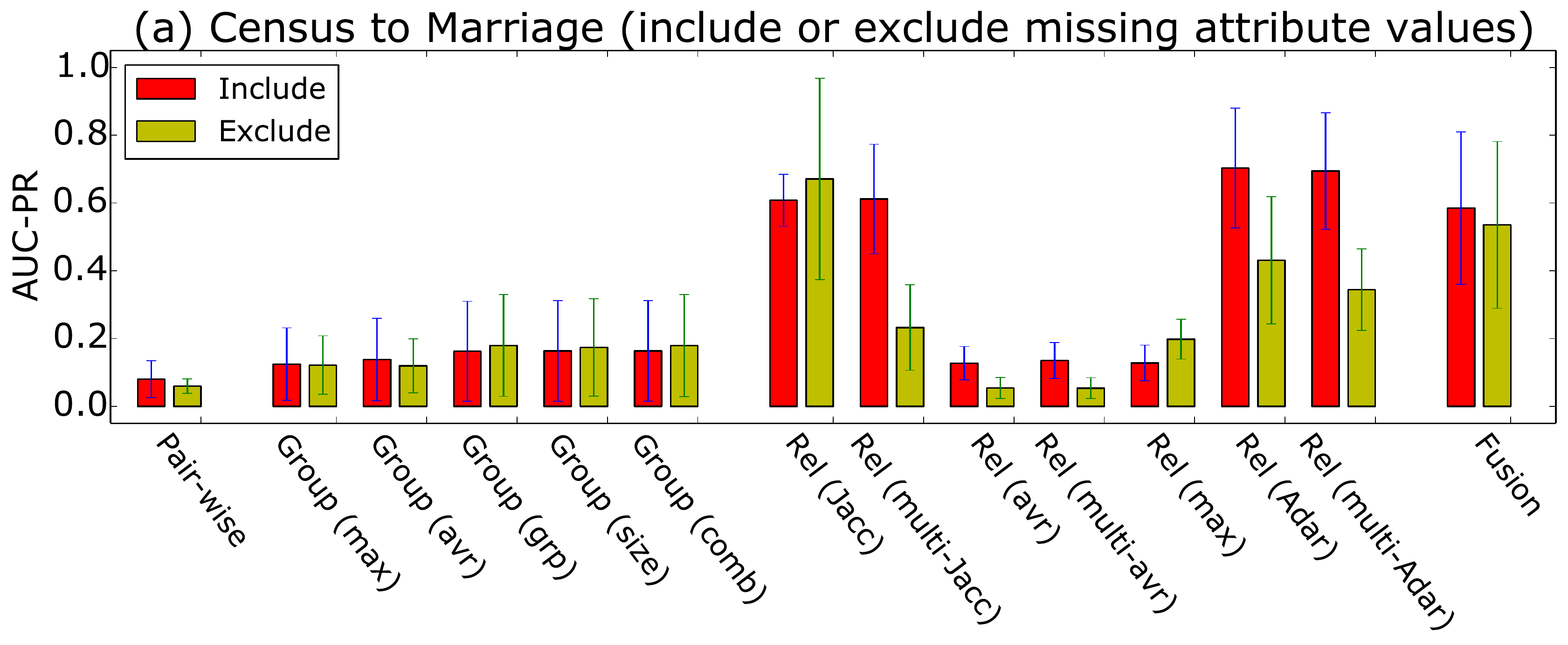}~~~~
  \includegraphics[width=0.49\textwidth]
                  {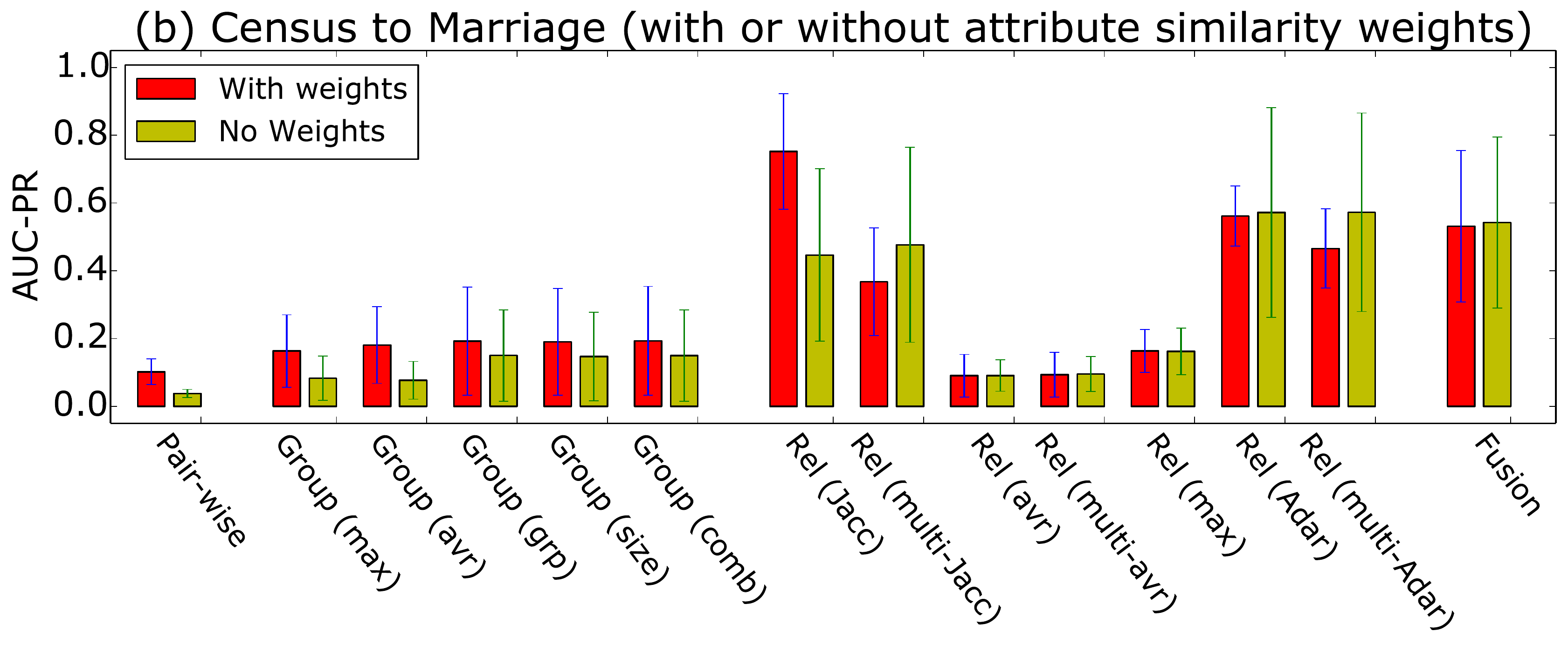}
  ~ \\[-2mm]
  \includegraphics[width=0.49\textwidth]
                  {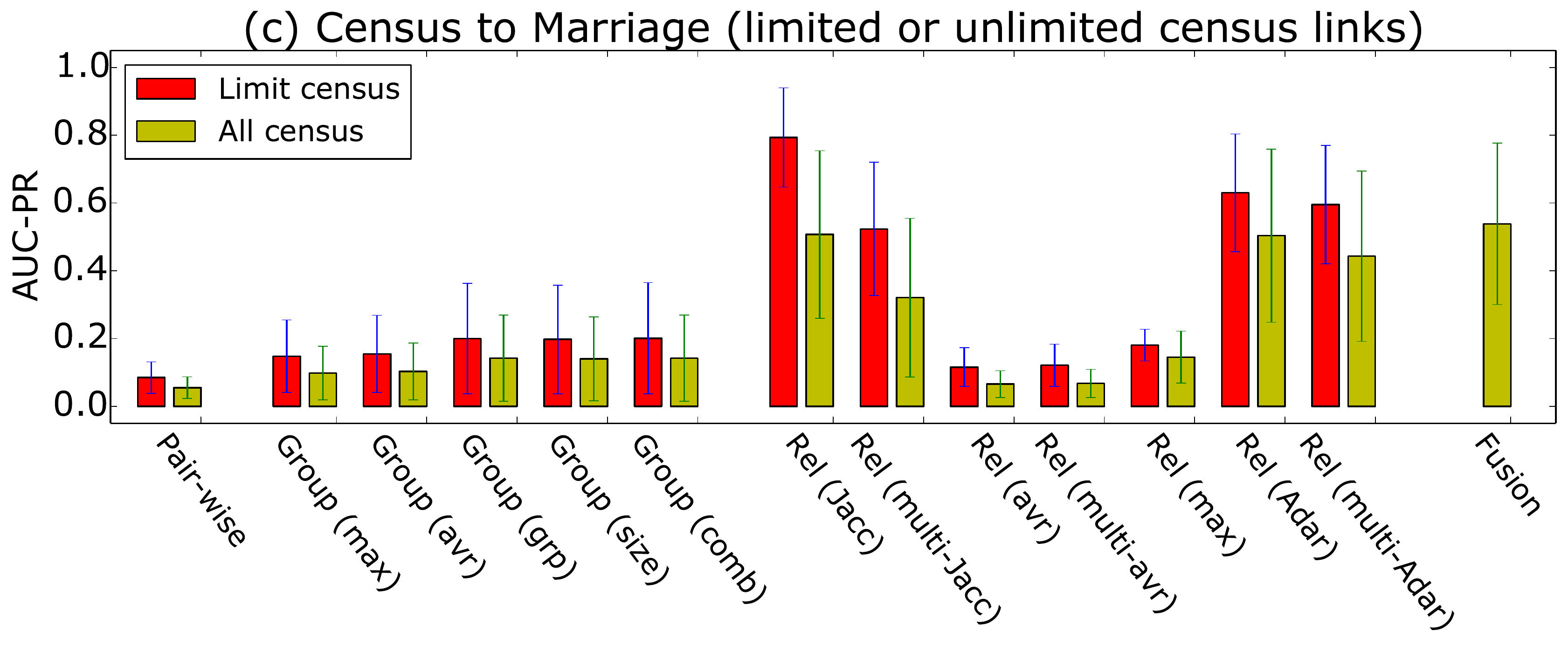}~~~~
  \includegraphics[width=0.49\textwidth]
                  {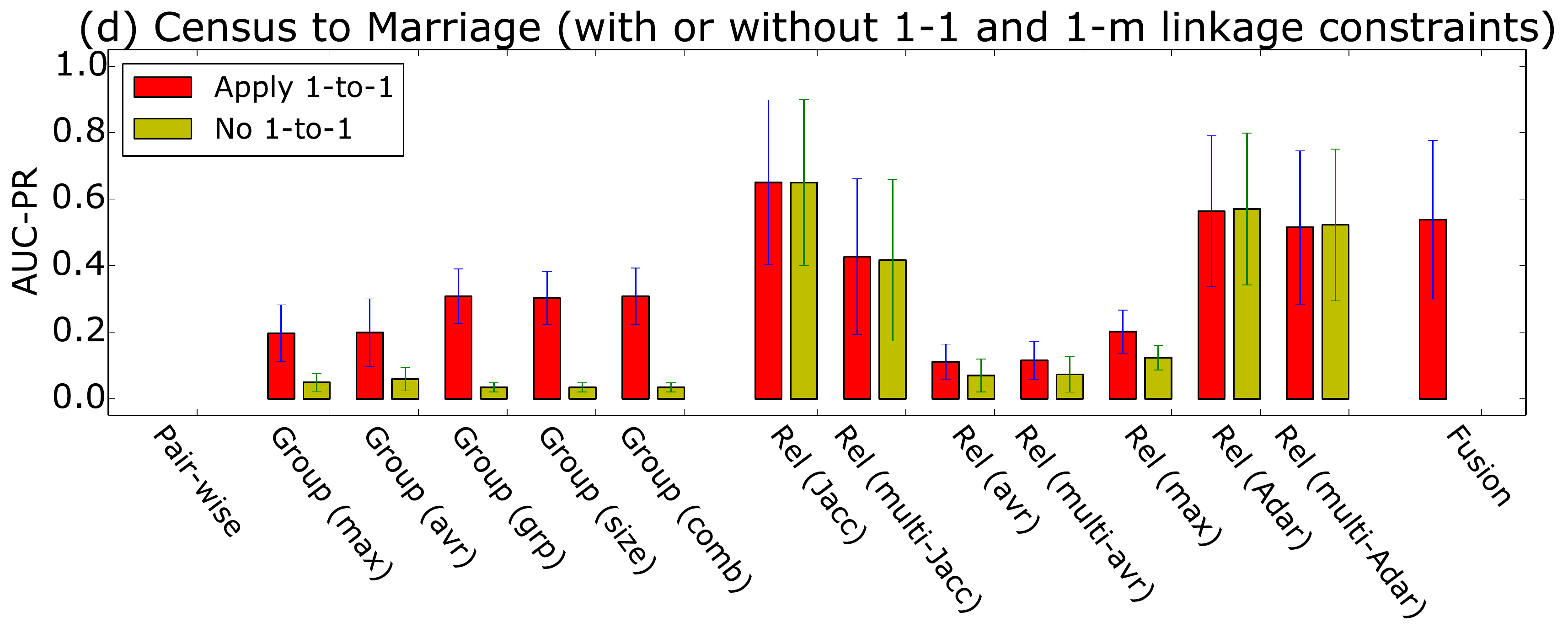}
  \caption{Area under the precision-recall curve (AUC-PR) results
           for census to marriage certificate pairs and different
           linkage options as discussed in Sect.~\ref{sec-data}.
           \label{fig-res-c-m}}
\end{figure*}

\begin{figure*}[t!]
  \centering
  \includegraphics[width=0.49\textwidth]
                  {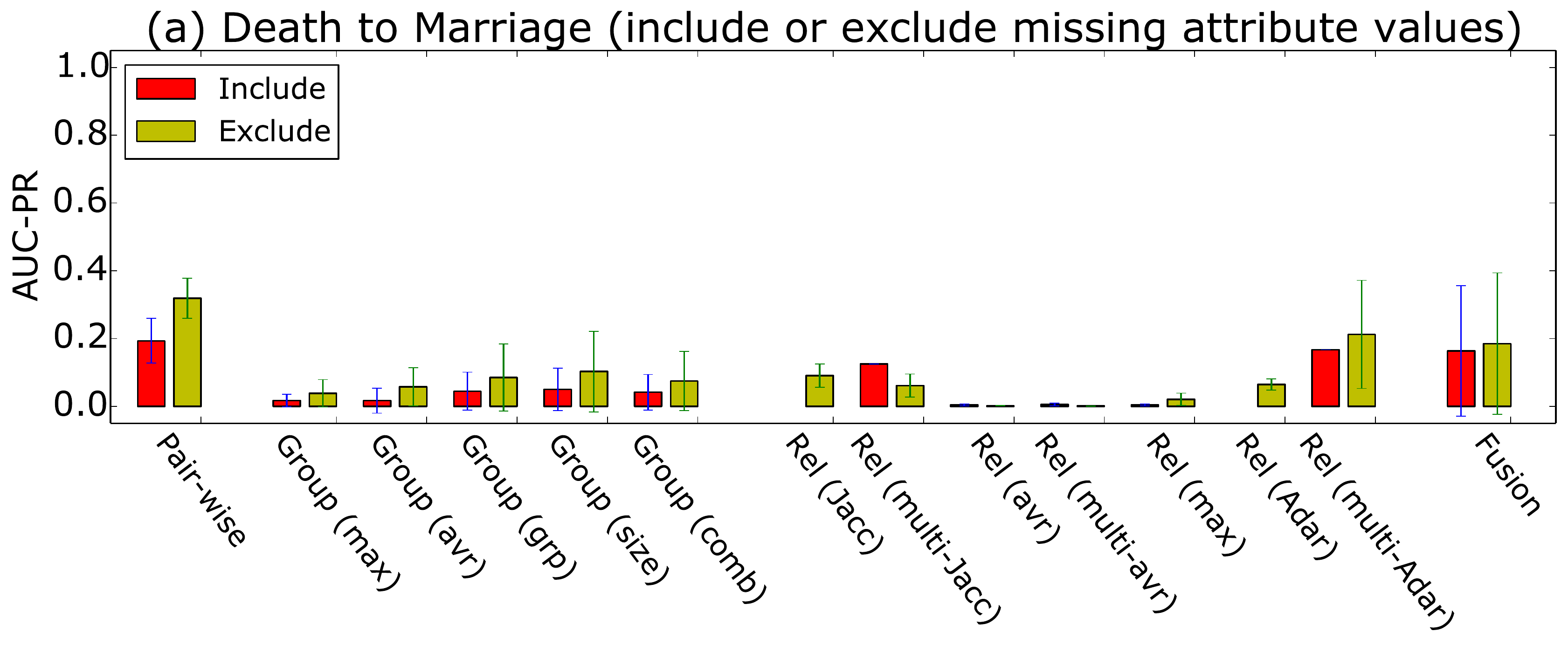}~~
  \includegraphics[width=0.49\textwidth]
                  {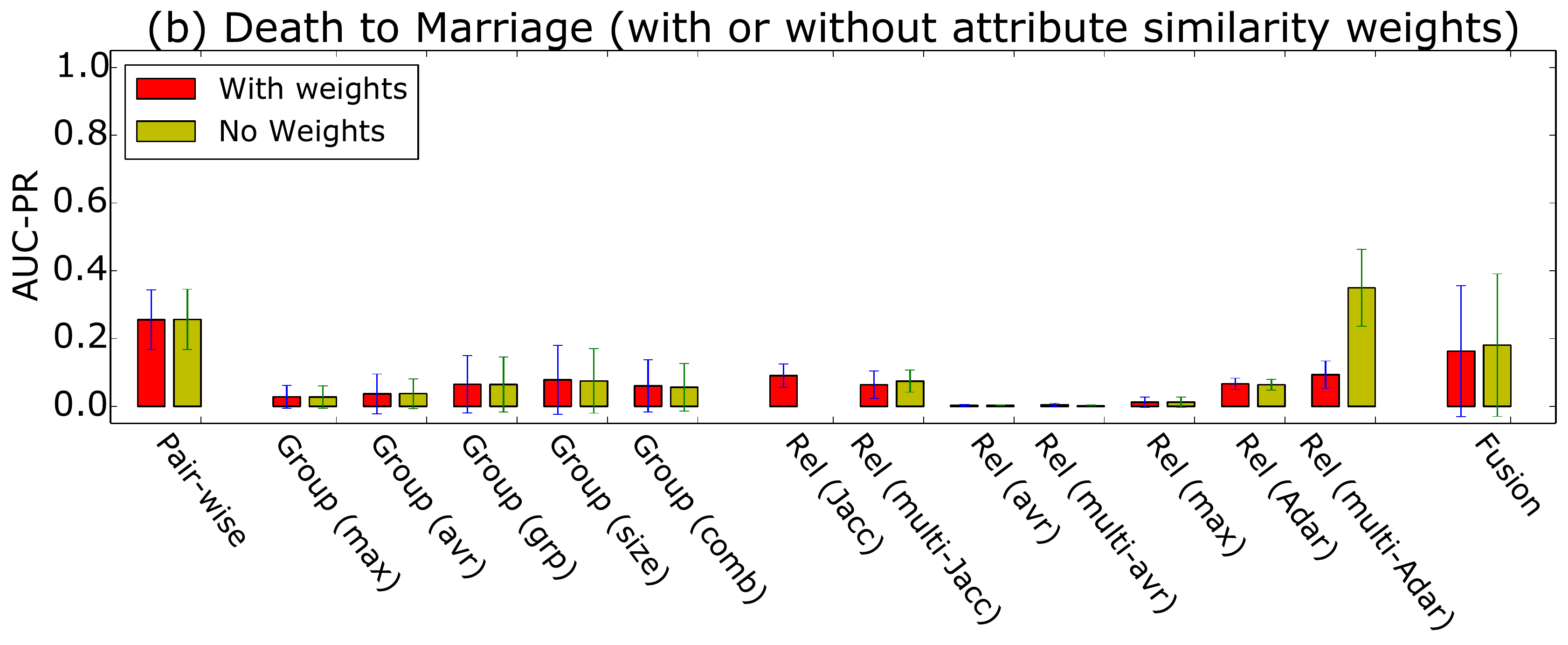}
  ~ \\[-2mm]
  \includegraphics[width=0.49\textwidth]
                  {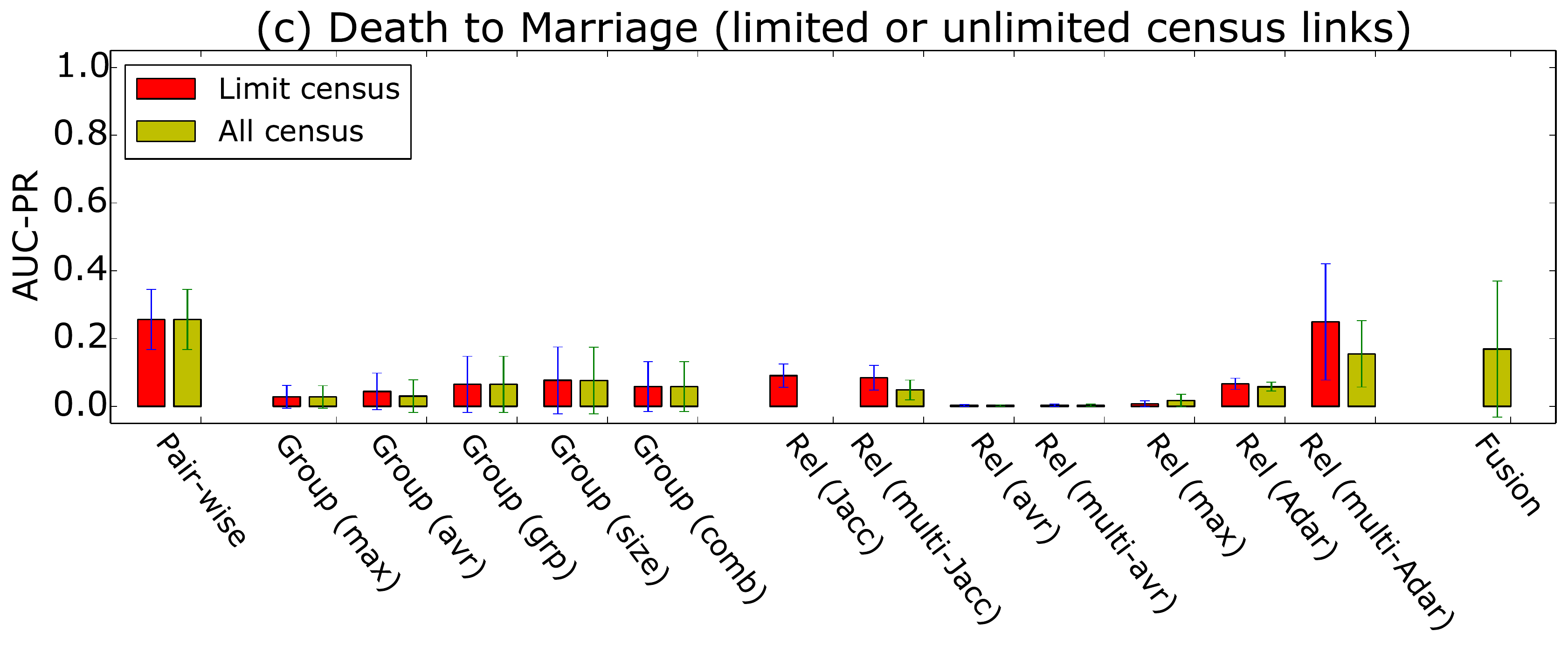}~~
  \includegraphics[width=0.49\textwidth]
                  {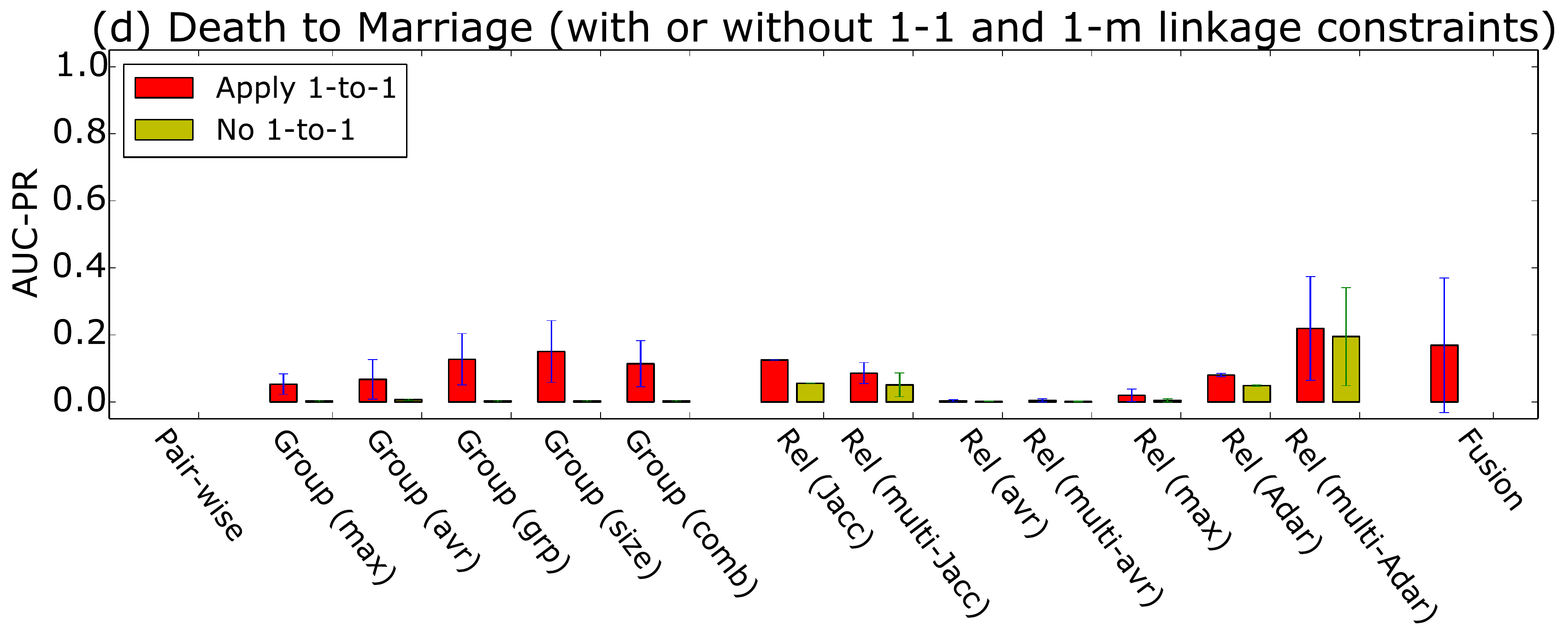}
  \caption{Area under the precision-recall curve (AUC-PR) results
           for death to marriage certificate pairs and different
           linkage options as discussed in Sect.~\ref{sec-data}.
           \label{fig-res-d-m}}
\end{figure*}

\begin{figure*}[t!]
  \centering
  \includegraphics[width=0.49\textwidth]
                  {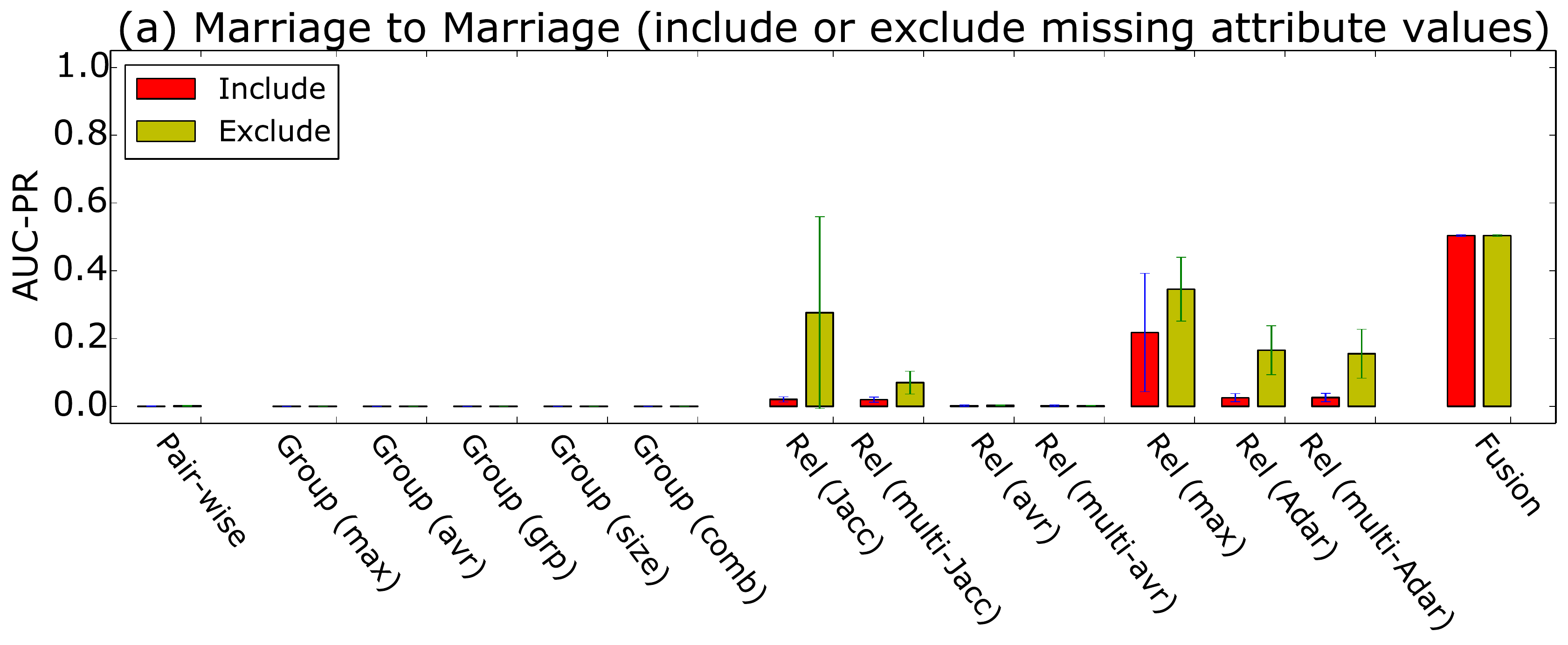}~~
  \includegraphics[width=0.49\textwidth]
                  {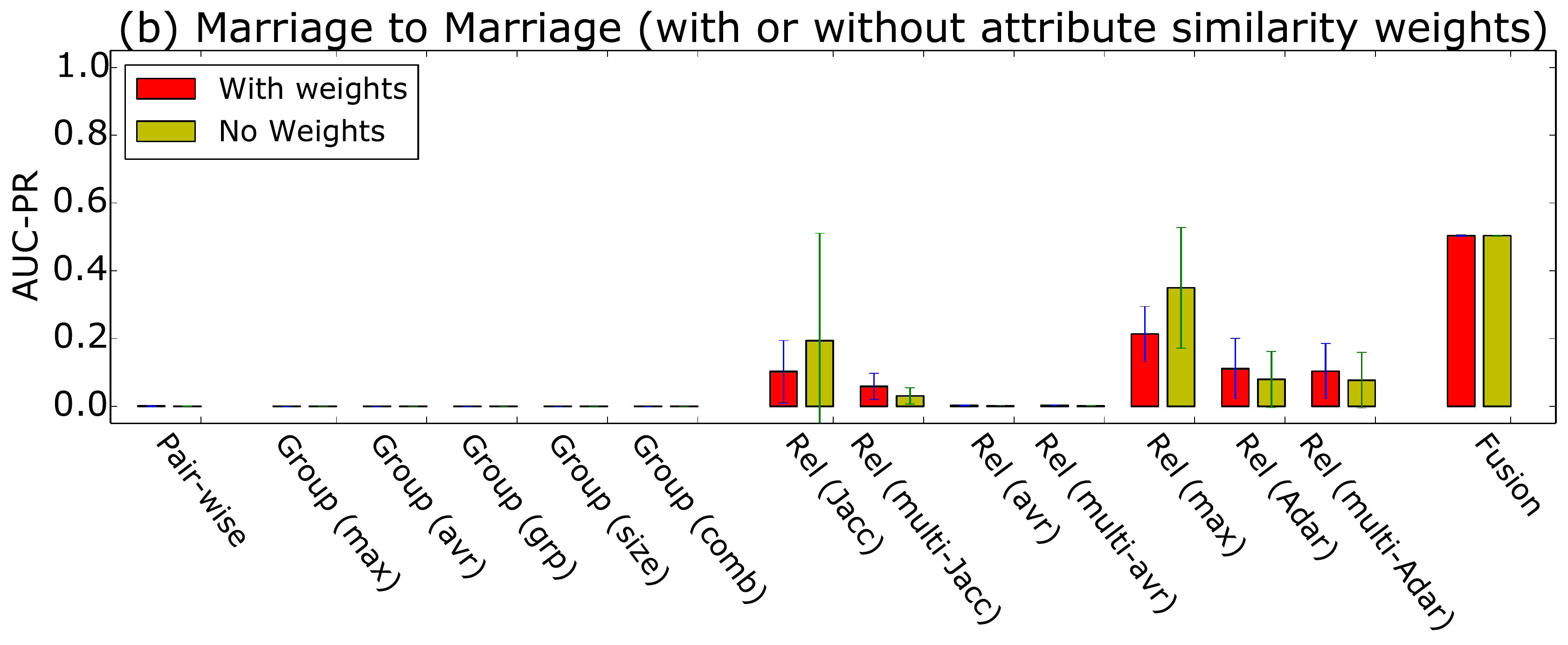}
  ~ \\[-2mm]
  \includegraphics[width=0.49\textwidth]
                  {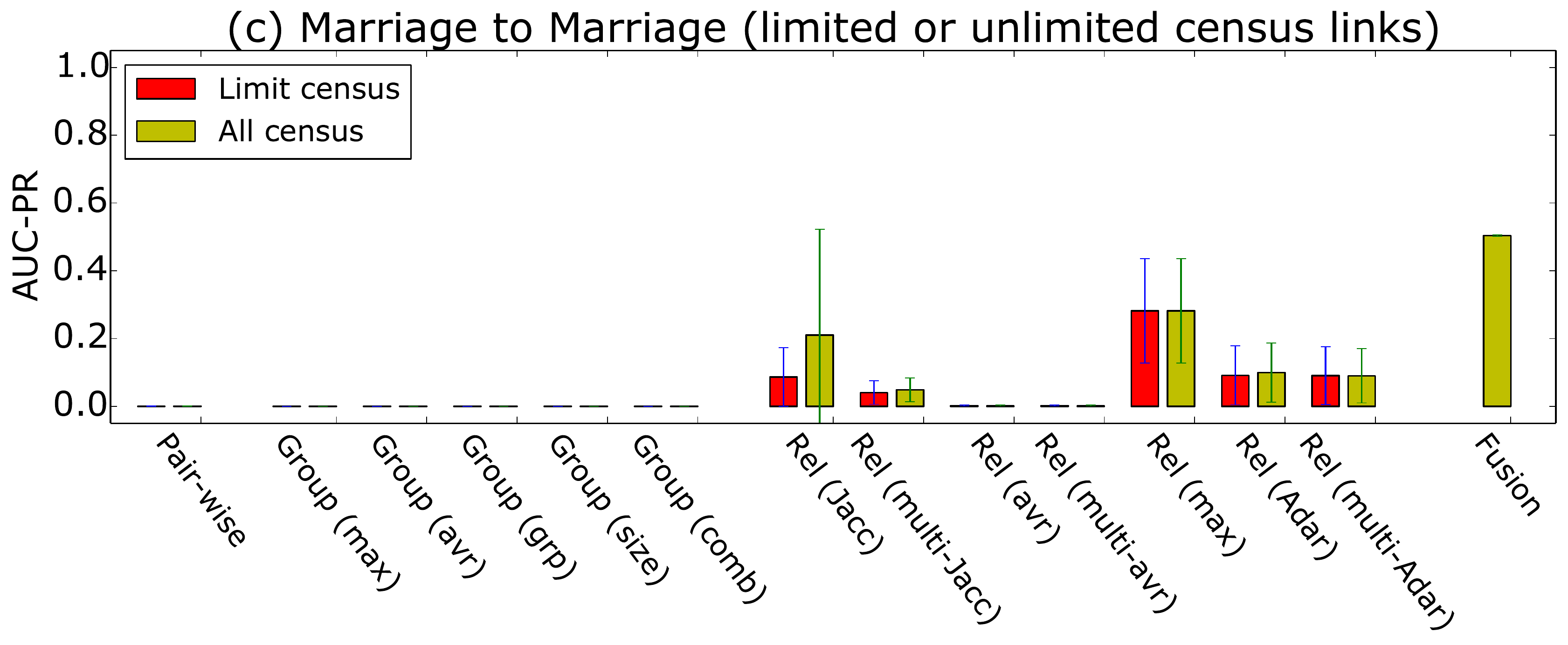}~~
  \includegraphics[width=0.49\textwidth]
                  {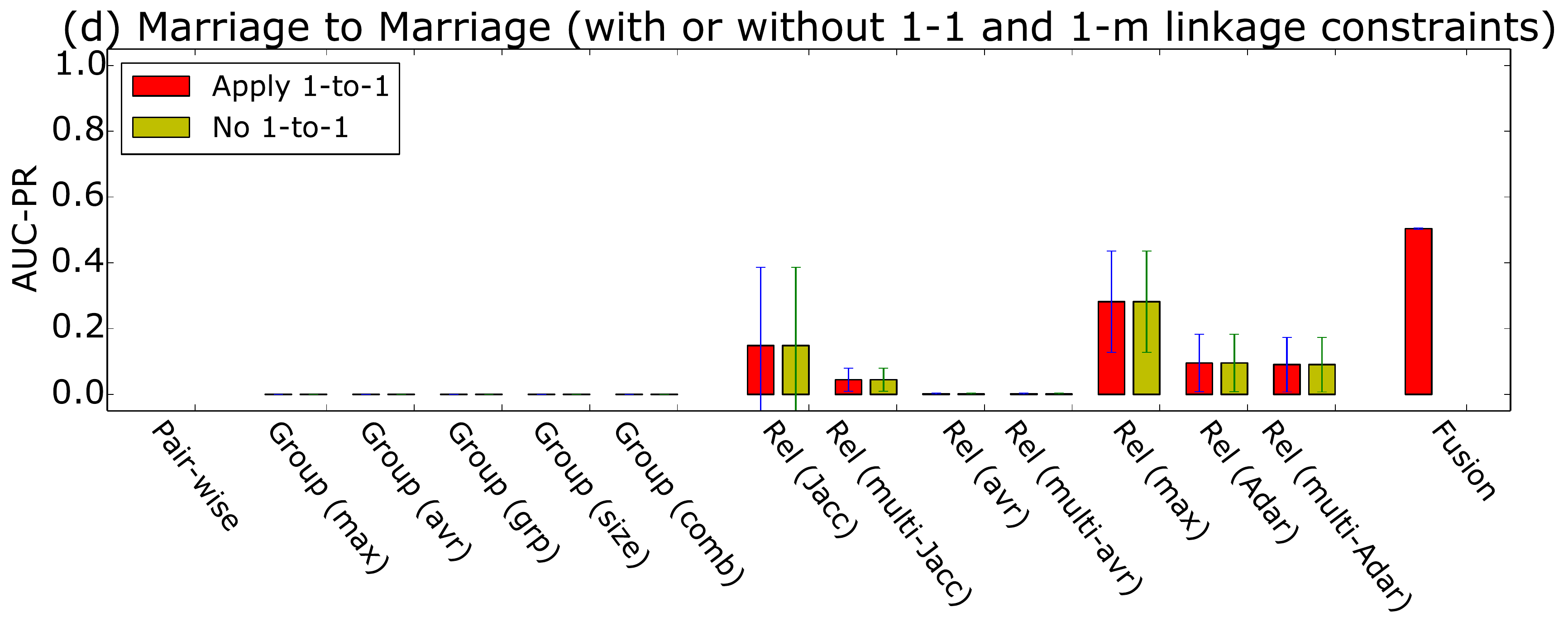}
  \caption{Area under the precision-recall curve (AUC-PR) results
           for marriage to marriage certificate pairs and different
           linkage options as discussed in Sect.~\ref{sec-data}.
           \label{fig-res-m-m}}
\end{figure*}

We report our findings in Fig.~\ref{fig-res-all} for all types of
certificate pairs, and in Figs.~\ref{fig-res-b-c}
to~\ref{fig-res-m-m} for individual certificate type pairs. We show
average and standard deviations of AUC-PR results as run over the
different settings of the options described above.

As all result figures show, the pair-wise linkage of individuals alone
does not lead to acceptable linkage quality at all. In fact, the
highest AUC-PR values obtained with pair-wise linkage only is around
$0.4$ for census to census certificate pairs, indicating that manual
links are mainly based on more than just attribute similarities.

Similarly, applying a relational or group linkage approach
individually only improves linkage quality to a certain degree. Fusion
of the linked certificate pairs obtained by both relational and group
linkage provides the best results for all types of pairs. With regard
to relational similarities, somewhat surprisingly the Jaccard
similarity achieved the best results overall, outperforming more
advanced relational similarity methods. For group linkage, no clear
winner can be found for the different group similarities.

With regard to the different linkage options we investigated,
including missing attribute values seems to lead to slightly better
results than excluding them; using attribute weights in the individual
similarity calculation does surprisingly not lead to any improvement;
and limiting census links to the closest temporal census certificates
can improve linkage quality. Finally, applying 1-to-1 and 1-to-many
constraints does significantly improve the quality of group linkage,
however it does not lead to improvements for relational linkage
methods.

Overall, as the results show, the overlap between the links obtained
with our automatic approach compared to the links manually
identified by domain experts is below what we expected with
state-of-the-art advanced linkage techniques that have been able to
achieve very high linkage quality on bibliographic
databases~\cite{Bha07,Don05}. We did not expect to achieve very high
AUC-PR of $0.9$ or above in the first place, given the challenging
nature of the data used in the experiments and the extensive manual
linkage conducted by experienced domain experts~\cite{New11,Rei02}.
However, it seems advanced linkage techniques developed for
bibliographic data cannot be directly used for linking complex
personal data, such as the ones we aim to link in our work. More
development of new linkage techniques that better exploit the
contents, roles, temporal aspects, and relationships between
individuals and groups, is therefore required.


\section{Conclusions and Future Work}
\label{sec-concl}

We have presented a novel approach to link complex personal data from
(historical) birth, death, marriage, and census certificates. Our
approach combines individual, group, and relational linkage methods,
and incorporates temporal, as well as 1-to-1 and 1-to-many linkage
constraints. An experimental evaluation on real Scottish data
highlighted the challenges of linking such complex data, where no
approach achieved high linkage quality compared to a careful manual
linkage as conducted by domain experts~\cite{Rei02}. For these
specific data sets, we see the main reason for this to be the large
number of intrinsically difficult to link cases that were manually
linked by the domain experts despite not having high attribute,
relational, or group similarities; or having highly ambiguous name or
address values.

As future work we aim to improve each step of our automatic linkage
approach, starting with pair-wise similarity calculations that are
suitable for the highly skewed frequency distributions of attribute
values. We plan to explore different blocking techniques, such as the
sorted neighborhood method~\cite{Chr12}, to improve the scalability of
our approach, and we aim to incorporate a fully relational collective
entity resolution approach~\cite{Bha07} and evaluate how this can
improve linkage quality.

While currently the role pair types used to limit pair-wise individual
comparisons are set manually based on domain expertise, we plan to
investigate if association or pattern mining algorithms~\cite{Hip00}
can help to automatically identify role pairs that are likely to
correspond to true changes of people's roles over time. Finally, we
plan to refine the temporal linkage constraints by calculating
fine-tuned temporal similarity adjustments~\cite{Chr13,Li11b}. Our
overall goal is to develop highly accurate, scalable, and automatic
techniques for linking large-scale complex population databases.


\section*{Acknowledgements}

This work was partially supported by a grant from the Simons
Foundation, and partially funded by the Australian Research Council
under Discovery Project DP160101934, the Administrative Data Research
Centre Scotland (ADRC-Scotland), and the Digitising Scotland project.
The author would like to thank the Isaac Newton Institute for
Mathematical Sciences, Cambridge, for support and hospitality during
the programme Data Linkage and Anonymisation where parts of this work
was conducted (EPSRC grant EP/K032208/1).


\bibliographystyle{acm}

\begin{small}
\bibliography{paper}

\begin{thebibliography}{10}

\bibitem{Ant14a}
{\sc Antonie, L., Inwood, K., Lizotte, D.~J., and Ross, J.~A.}
\newblock Tracking people over time in 19th century {Canada} for longitudinal
  analysis.
\newblock {\em Machine Learning 95\/} (2014), 129--146.

\bibitem{Bha07}
{\sc Bhattacharya, I., and Getoor, L.}
\newblock Collective entity resolution in relational data.
\newblock {\em ACM TKDD 1}, 1 (2007).

\bibitem{Blo15}
{\sc Bloothooft, G., Christen, P., Mandemakers, K., and Schraagen, M.}
\newblock {\em Population Reconstruction}.
\newblock Springer, 2015.

\bibitem{Chi14}
{\sc Chiang, Y.-H., Doan, A., and Naughton, J.~F.}
\newblock Tracking entities in the dynamic world: A fast algorithm for matching
  temporal records.
\newblock {\em PVLDB 7}, 6 (2014).

\bibitem{Chr09b}
{\sc Christen, P.}
\newblock Development and user experiences of an open source data cleaning,
  deduplication and record linkage system.
\newblock {\em SIGKDD Explorations\/} (2009).

\bibitem{Chr12}
{\sc Christen, P.}
\newblock {\em Data Matching -- Concepts and Techniques for Record Linkage,
  Entity Resolution, and Duplicate Detection}.
\newblock Springer, 2012.

\bibitem{Chr13}
{\sc Christen, P., and Gayler, R.~W.}
\newblock Adaptive temporal entity resolution on dynamic databases.
\newblock In {\em PAKDD\/} (Gold Coast, Australia, 2013).

\bibitem{Dav06}
{\sc Davis, J., and Goadrich, M.}
\newblock The relationship between {Precision-Recall} and {ROC} curves.
\newblock In {\em ACM ICML\/} (Pittsburgh, 2006), pp.~233--240.

\bibitem{Don05}
{\sc Dong, X.~L., Halevy, A., and Madhavan, J.}
\newblock Reference reconciliation in complex information spaces.
\newblock In {\em SIGMOD\/} (Baltimore, 2005).

\bibitem{Fu14b}
{\sc Fu, Z., Christen, P., and Zhou, J.}
\newblock A graph matching method for historical census household linkage.
\newblock In {\em PAKDD\/} (Tainan, Taiwan, 2014).

\bibitem{Fu12}
{\sc Fu, Z., Zhou, J., Christen, P., and Boot, M.}
\newblock Multiple instance learning for group record linkage.
\newblock In {\em PAKDD\/} (Kuala Lumpur, 2012).

\bibitem{Her07}
{\sc Herzog, T.~N., Scheuren, F.~J., and Winkler, W.~E.}
\newblock {\em Data quality and record linkage techniques}.
\newblock Springer, 2007.

\bibitem{Hip00}
{\sc Hipp, J., G\"{u}ntzer, U., and Nakhaeizadeh, G.}
\newblock Algorithms for association rule mining -- a general survey and
  comparison.
\newblock {\em SIGKDD Explorations 2}, 1 (2000), 58--64.

\bibitem{Kop10b}
{\sc K{\"o}pcke, H., Thor, A., and Rahm, E.}
\newblock Evaluation of entity resolution approaches on real-world match
  problems.
\newblock {\em PVLDB 3}, 1-2 (2010), 484--493.

\bibitem{Kum14}
{\sc Kum, H.-C., Krishnamurthy, A., Machanavajjhala, A., and Ahalt, S.~C.}
\newblock Social genome: Putting big data to work for population informatics.
\newblock {\em IEEE Computer 47}, 1 (2014).

\bibitem{Li11b}
{\sc Li, P., Dong, X.~L., Maurino, A., and Srivastava, D.}
\newblock Linking temporal records.
\newblock {\em PVLDB 4}, 11 (2011).

\bibitem{Nau10}
{\sc Naumann, F., and Herschel, M.}
\newblock {\em An introduction to duplicate detection}.
\newblock Synthesis Lectures on Data Management. Morgan and Claypool
  Publishers, 2010.

\bibitem{New11}
{\sc Newton, G.}
\newblock Recent developments in making family reconstitutions.
\newblock {\em Local Population Studies 87}, 1 (2011), 84--89.

\bibitem{On07}
{\sc On, B.-W., Koudas, N., Lee, D., and Srivastava, D.}
\newblock Group linkage.
\newblock In {\em IEEE ICDE\/} (Istanbul, 2007).

\bibitem{Ped11}
{\sc Pedregosa, F., Varoquaux, G., Gramfort, A., Michel, V., Thirion, B.,
  et~al.}
\newblock Scikit-learn: Machine learning in {Python}.
\newblock {\em The Journal of Machine Learning Research 12\/} (2011),
  2825--2830.

\bibitem{Rei02}
{\sc Reid, A., Davies, R., and Garrett, E.}
\newblock Nineteenth-century {Scottish} demography from linked censuses and
  civil registers.
\newblock {\em History and Computing 14}, 1-2 (2002).

\bibitem{Wri73}
{\sc Wrigley, E.~A., and Schofield, R.~S.}
\newblock Nominal record linkage by computer and the logic of family
  reconstitution.
\newblock {\em Identifying people in the past\/} (1973), 64--101.

\end{thebibliography}
\end{small}

\end{document}